\documentstyle[prl,aps,multicol]{revtex}
\input{epsf}
\begin{document}
\setcounter{page}{1}
\draft
\title{\large {\bf Chaotic enhancement in microwave ionization  
of Rydberg atoms}} 
\author{Giuliano Benenti$^{(a,b,c,+)}$,
Giulio Casati$^{(a,b,c)}$, and
Dima L. Shepelyansky$^{(d,*)}$}
\address{$^{(a)}$International Centre for the Study of Dynamical
Systems,}
\address{Universit\`a
di Milano, Sede di Como, Via Lucini 3, 22100
Como, Italy}
\address{$^{(b)}$Istituto Nazionale di Fisica della Materia,
Unit\`a di Milano, Via Celoria 16, 20133 Milano, Italy}
\address{$^{(c)}$INFN, Sezione di Milano, Via Celoria 16, 20133 Milano,
Italy}
\address{$^{(d)}$Laboratoire de Physique Quantique, UMR C5626 du CNRS, 
Universit\'e Paul Sabatier, 31062,
Toulouse, France}
\date{\today}
\maketitle
\begin{abstract}
The microwave ionization of internally chaotic Rydberg atoms is
studied analytically and numerically. The internal chaos is induced
by magnetic or static electric fields. This leads to a
chaotic enhancement of microwave excitation.
The dynamical localization theory  gives a detailed
description of the excitation process even in a regime where up to few
thousands photons are required to ionize one atom.
Possible laboratory experiments are also discussed.
\end{abstract}

\begin{multicols}{2}
\narrowtext

\section{Introduction}
The pioneering experiment of Bayfield and Koch performed in 1974 \cite{BK}
attracted a great  interest to ionization of highly excited
hydrogen and Rydberg atoms in a microwave field 
\cite{DKS,CCSG87,Uzy,IEEE,JRP,KRP}. 
The main reason of this
interest is due to the fact that such ionization
requires the absorption of a large number of  photons (about $20-70$)
and can be explained only as a result of the appearence of dynamical chaos and 
diffusive energy excitation in the corresponding
classical system. Indeed the critical border $\epsilon_c$ for the microwave field intensity above which classical chaotic motion takes place is given by \cite{DKS}:
\begin{eqnarray}
\label{chbdr} 
\epsilon_0 > \epsilon_c \approx \frac{1}{49 {\omega_0}^{1/3}}.
\end{eqnarray}
Here,  $\epsilon$ and $\omega$ are the microwave field strength and frequency, $\epsilon_0 = \epsilon n_0^4$ and $\omega_0 = \omega n_0^3$
are the  rescaled values, and $E_0 = - 1/2n_0^2$
is the initial unperturbed energy of the atom 
(see rescaling details below; atomic units are used).
It is also assumed that $\omega_0 \geq 1$ and that the initial orbital momentum
$l < (3/\omega)^{1/3}$.
For $\epsilon_0 < \epsilon_c$ the  electron energy $E$
only performs small oscillations  around its initial value and therefore
 ionization is impossible in the classical system.
Above the chaos border instead, the electron's energy 
increases in a diffusive way with a diffusion rate per unit time
given by \cite{DKS}:
\begin{eqnarray}
\label{chdif} 
D_{E} = \frac{(\Delta E)^2}{\Delta t} \approx
0.5 \frac{\epsilon^2}{\omega^{4/3} n_0^3}.
\end{eqnarray}
This diffusion leads to  electron's  ionization after 
a typical diffusive time scale $t_D \sim E_0^2/D_E$, with
$E_0 = - 1/2n_0^2$ being the initial energy. Such classical diffusive ionization
requires many microwave periods ($t_D \omega/2 \pi \gg 1$)
and quantum interference effects can suppress this diffusion
leading to quantum localization of chaos \cite{CCSG87,IEEE}.
Such dynamical localization of chaos leads to a quantum
probability distribution $f_N$ exponentially localized in the number of absorbed 
photons $N_\phi=(E-E_0)/\omega$, namely  
$f_N \propto \exp(-2 |N_\phi|/\ell_{\phi_\omega})$. For the general 
case of monochromatic field excitation in a complex spectrum
the localization
length $\ell_{\phi_\omega}$, measured in the number of photons,
can be determined via the one-photon transition rate $\Gamma \sim D_E/\omega^2$
and the density of coupled states $\rho_c$ \cite{Dima87}:
\begin{eqnarray}
\label{lphi} 
\ell_{\phi_\omega}= 2\pi \Gamma \rho_c = 
3.33 \frac{\epsilon^2}{\omega^{10/3}}.
\end{eqnarray}
The last equality in the above equation corresponds to the hydrogen case, where
$\rho_c=n_0^3$ due to Coulomb degeneracy (which, as is known, is responsible for 
the appearence of an additional integral on motion \cite{IEEE}).
Quite obviously, in case of strong localization, namely when the localization 
length is less than the number of photons
required for ionization, $N_I = |E_0|/\omega = n_0/2\omega_0$, 
the quantum ionization is exponentially small and therefore negligible compared to its
classical value. In the opposite case $\ell_{\phi_\omega} > N_I$,
namely  above the quantum delocalization border $\epsilon_{q0}$
\begin{eqnarray}
\label{delocb} 
\epsilon_0>\epsilon_{q0} =\frac{\omega_0^{7/6}}{\sqrt{6.6 n_0}},
\end{eqnarray}
 quantum ionization takes place in close to the classical case \cite{IEEE}.
We note that the above dynamical localization represents a deterministic 
analog of the Anderson localization in disordered quasi--one--dimensional
chains. In our case the site index corresponds to the photon number $N_\phi$,
while $N_I$ plays the role of the effective sample size.

The above theoretical results 
have been checked by different  groups \cite{Uzy,JRP} and 
more recently reconfirmed in numerical simulations with 
newly developed algorithms \cite{BD95,Kubo}.
The predictions of dynamical localization have been also confirmed
by laboratory experiments \cite{K1988,B1989,Walther}.
However,  in the mesoscopic solid-state language,
the  effective ``sample size'' available in these experiments, corresponding 
to $N_I \approx 20$
was too short to test with sufficient accuracy the theoretical predictions.
In fact in this situation one observes significant mesoscopic
fluctuations in the ionization border which have been discussed in detail in
\cite{JRP,KRP}. While the main structure of these mesoscopic
fluctuations can be well described by the quantum Kepler map
\cite{Brasil} it is still highly desirable to have much longer
``samples'' with $N_I \geq 100$ to study experimentally
the dynamical localization of chaos in more detail. We note that recently
the dynamical localization has been also observed in experiments with 
cold atoms propagation
\cite{Raizen}; however, in this situation there are other
experimental restrictions.

In order to have larger  $N_I$ values, one has, either  to increase $n_0$
or to take $\omega_0 \ll 1$. However, the first possibility
is restricted by experimental conditions where one prefers to have
$n_0 < 100$. The second choice leads to a regime close to the  static field
ionization in which no classical chaos exists; in any case, localization of classically chaotic motion does not take place for $\omega_0 <<1$.
In order to actually  have  large $N_I$ values,  it is necessary to take a different
approach and  work
with atoms which are classically chaotic already in the absence of
the microwave field. There are two main possibilities to have 
chaotic Rydberg atoms. One way is to put hydrogen or Rydberg atoms
in a magnetic field. In this case, the classical dynamics becomes chaotic
when the Larmor frequency is comparable with the 
unperturbed Kepler frequency \cite{Delande,Wintgen,Robnik}. 
Another possibility
is to use Rydberg atoms with quantum defects in a static electric field.
Recent investigations have shown that for a sufficiently strong 
static field the level spacing statistics is similar to the 
case of random matrix theory (RMT) \cite{Kleppner}. From the experimental
viewpoint the case with a static electric field is 
simpler to deal with and in fact can be studied
in laboratory experiments similar to \cite{Gall1,Gall2,Klep2}.
For such atoms the chaos border for the microwave field 
drops to zero and therefore
even at very small microwave field one can expect to see diffusive excitation
in energy. In addition, such a diffusion can take place for 
much lower frequencies with $1/n_0 \ll \omega_0 \ll 1$, and this fact 
allows to increase the values of $N_I$ up to few thousands.

The  investigation of the interaction of chaotic Rydberg atoms with a microwave
radiation is also interesting from another viewpoint. Indeed, 
it is known that a chaotic structure of eigenstates leads to
a strong enhancement of the interaction. In nuclear physics, as was shown by
Sushkov and Flambaum \cite{Sushkov-Flambaum},  this effect leads to an
enhancement of weak interaction and parity violation
by a factor of thousand or more.
Also in the problem of Anderson localization in disordered
solid--state systems 
it has been found that a short range repulsive/attractive
interaction between two particles
can strongly enhance their propagation \cite{Dima94}. All this indicates
that a chaotic structure of Rydberg atoms can strongly increase
their interaction with radiation. This should lead 
to a significant decrease of the quantum delocalization border
as compared to the standard case of internally non-chaotic atoms
studied in \cite{CCSG87,IEEE}. In fact, 
the experiments by Gallagher {\it et al.}
\cite{Gall1,Gall2} indicated a lower ionization border than for 
 the hydrogen case \cite{KRP}. The physical mechanism of such
ionization, proposed by Gallagher {\it et al.} \cite{Gall1,Gall2},
is based  on a picture of successive Landau-Zener crossings in a slowly
oscillating microwave field. However, the question how 
such transitions can proceed to high levels was never studied
in detail. Moreover, for $\omega > 1/n^4$ one enters in a 
multiphoton regime which was not discussed in \cite{Gall1,Gall2}.

Due to the above reasons,  the investigation of chaotic enhancement 
of microwave ionization of Rydberg atoms allows to address a new
physical regime in which thousands of photons are required to 
ionize one atom. Our theoretical and numerical results
indeed clearly demostrate the existence of such enhancement
and provide a clear physical picture of the ionization process.
In section II we discuss the case of atoms in parallel
magnetic and microwave fields, while the case of Rydberg
atoms in static and microwave fields is analysed in Section III.
The main results are discussed in Section IV. Some  results have been
presented in \cite{mag1,mag2,qdef}.

\section{The hydrogen atom in magnetic and microwave fields}
The Hamiltonian of a hydrogen atom in parallel microwave and 
uniform magnetic fields writes 
\begin{eqnarray}
\label{ham1}
\begin{array}{c}
\displaystyle{H=\frac{p_\rho^2}{2}+\frac{p_z^2}{2}+\frac{L_z^2}{2\rho^2}
-\frac{1}{\sqrt{\rho^2+z^2}}+\frac{\omega_L L_z}{2}+}\cr\cr
\displaystyle{
\frac{\omega_L^2}{8}\,\rho^2+\epsilon z \sin(\omega t)},
\end{array}
\end{eqnarray} 
where $z$ is the direction of the fields, $\rho=(x^2+y^2)^{1/2}$,
$\omega_L=B/c=B(T)/B_0$ is the cyclotron frequency, $B_0=2.35\times
10^5 T$, $\epsilon$ and $\omega$ are the microwave strength and frequency
respectively (atomic units are used). Due to the cylindrical symmetry, 
the $z$ component of the angular momentum $L_z$ is a constant of the  
motion and here we consider $L_z=0$.

The above Hamiltonian, expressed as a function of cartesian
coordinates  $\{x_i\}$, their conjugate momenta $\{p_i\}$, the
time $t$ and the parameters $L_z$, $\omega_L$, $\epsilon$, $\omega$
has the property
\begin{eqnarray}
\label{scaled}
\begin{array}{c}
\displaystyle{
\tilde{H}= H(\{\lambda x_i\},\{\lambda^{-1/2} p_i\},\lambda^{3/2} t,
\lambda^{1/2} L_z,\lambda^{-3/2}\omega_L,}\cr\cr
\displaystyle{
\lambda^{-2}\epsilon,\lambda^{-3/2}\omega)  
=\lambda^{-1}H(\{x_i\},\{p_i\},t,L_z,\omega_L,
\epsilon,\omega).}
\end{array}
\end{eqnarray}
If we choose $\lambda=1/n_0^2$, with $E_0=-1/2 n_0^2$ the initial energy,
the classical dynamics depends on $n_0$ only via the scaled variables
\cite{IEEE,Delande,Wintgen,Robnik} 
\begin{eqnarray}
\label{scalpar} 
L_z/n_0, \quad \omega_L n_0^3, \quad \epsilon_0=
\epsilon n_0^4,
\quad \omega_0=\omega n_0^3.
\end{eqnarray}
The coordinates and momenta scale as $\tilde{x}_i=x_i/n_0^2$,
$\tilde{p}_i=p_in_0$ and the rescaled time $\tilde{t}=t/n_0^3$,
up to a factor of $1/2\pi$, counts the number of Kepler
periods in the electron motion in the absence of external fields 
($\omega_L=\epsilon=0$). 

In order to study the dynamics of the time--dependent Hamiltonian
(\ref{ham1}), it is possible to introduce an extended phase space in
which the Hamiltonian becomes conservative with 
respect to a fictitious time $\eta$.
The new Hamiltonian writes
$K=\tilde{H}+\omega_0 \tilde{N}$, with
$\phi=\omega_0 \tilde{t}=\omega t$
and $\tilde{N}$ as new classical conjugate variables. 
The equations of motion for $\phi$ and $\tilde{N}$ are given by
\begin{eqnarray}
\label{floquet}
\frac{d\phi}{d\eta}=\frac{\partial K}{\partial \tilde{N}}=\omega_0,
\quad
\frac{d \tilde{N}}{d\eta}=-\frac{\partial K}{\partial \phi}=
-\frac{\partial \tilde{H}}{\partial \phi}.
\end{eqnarray}
Notice that $\eta$ is equal to the scaled time
$\tilde{t}$ up to an additive constant, and therefore the ordinary
Hamilton's equations follow for the other canonical variables. 
Since $K$ is a constant
of motion, in the quantum case the change of $\tilde{N}$ would give,
apart from a $1/n_0$ scaling factor, the
number of photons $\Delta N$ exchanged by the atom 
with the field \cite{IEEE}:
$\Delta \tilde{N}=-\Delta \tilde{H}/\omega_0=-(1/n_0)(\Delta E/\omega)=
-(1/n_0)\Delta N$.

\begin{figure}
\centerline{
\epsfxsize=8cm
\epsfysize=15cm
\epsffile{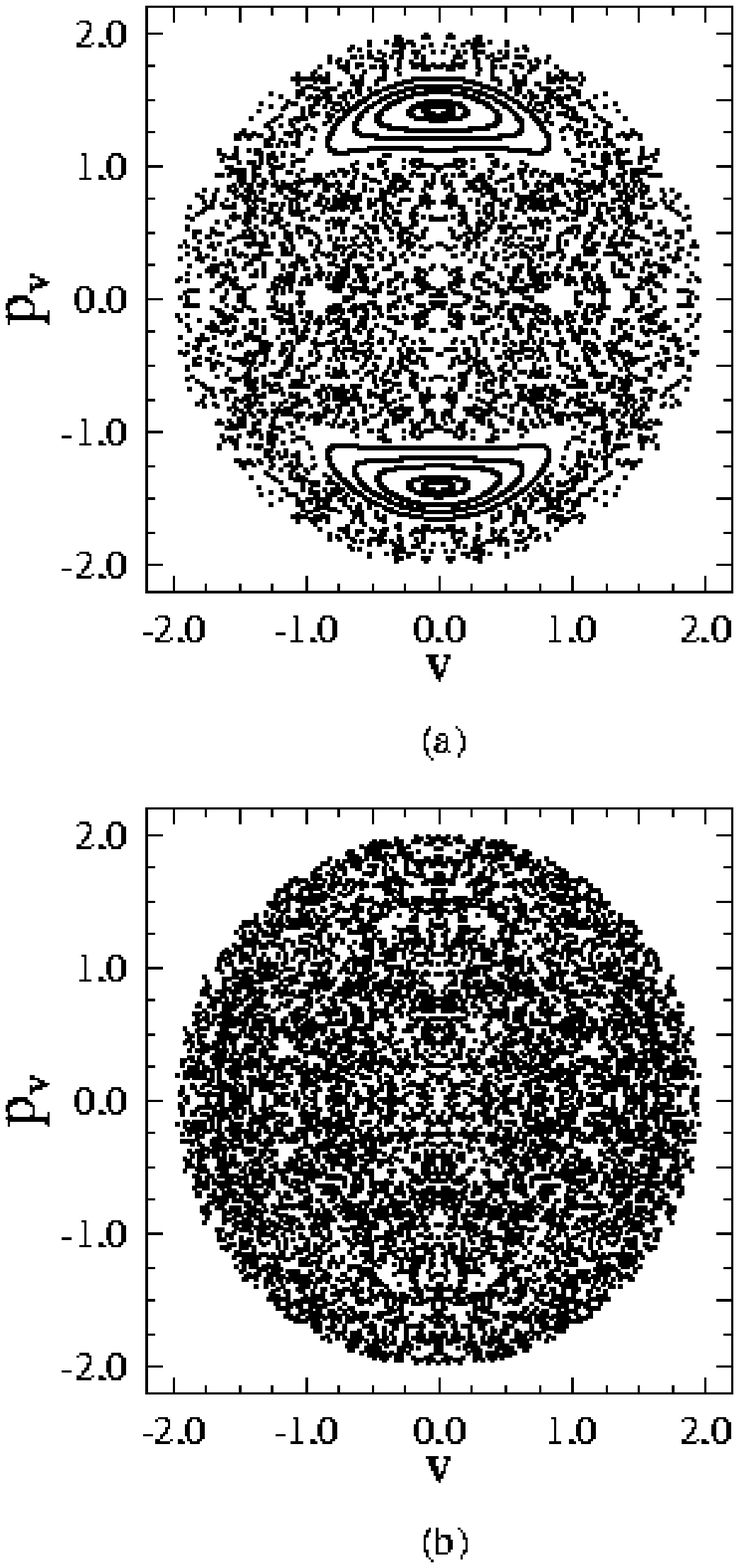}
}
\caption{
Poincar\'e surfaces of section at scaled energy 
$\tilde{E}=\tilde{H}(\epsilon=0)=-1/2$ and
(a) $\omega_L n_0^3=3$, (b)
$\omega_L n_0^3=9.2$. 
The sections are on the $v-p_v$ planes defined
by $u=0$, $p_u>0$. In (a) five regular and one chaotic trajectories
are shown, in (b) only one chaotic trajectory is followed.
The centre of the atom corresponds to $v=0$.
}
\label{fig1}
\end{figure}

The singularity 
of the Hamiltonian (\ref{ham1}) 
at $r=(\rho^2+z^2)^{1/2}=0$ 
can be removed by introducing the semi--parabolic coordinates
$u=(\tilde{r}+\tilde{z})^{1/2}$, $v=(\tilde{r}-\tilde{z})^{1/2}$ and 
the regularized time $\sigma$, defined as 
\cite{Delande,Wintgen,Robnik}
\begin{eqnarray}
\label{scaltau}
d\eta=d\tilde{t}=(u^2+v^2)\,d\sigma=2\tilde{r}d\sigma,
\end{eqnarray}
which changes faster than $\tilde{t}$ near the nucleus and more slowly 
far from it. 
The equations of motion
generated by the Hamiltonian (\ref{ham1}) are then equivalent
to the equations of motion generated, with respect to the new time 
$\sigma$, by the scaled and regularized 
Hamiltonian
\begin{eqnarray}
\label{kresc}
\cal{K}&  = (u^2+v^2) K.
\end{eqnarray}
Indeed, for any classical quantity $f(u,v,\phi;p_u,p_v,\tilde{N})$, we get
\begin{eqnarray}
\label{kpois}
\frac{df}{d\sigma}=\frac{df}{d\eta}\,\frac{d\eta}{d\sigma}=\{K,f\}(u^2+v^2),
\end{eqnarray}
where $\{K,f\}$ denotes the Poisson bracket between $K$ and $f$.
Thus, if we take $\tilde{N}_0=-(1/n_0)(E_0/\omega)$ as initial 
condition for $\tilde{N}$,
the compensated energy $K$ is equal to zero and we obtain
\begin{eqnarray}
\label{krpois}
\{\cal{K}& ,f\}=\{u^2+v^2,f\}K+(u^2+v^2)\{K,f\}=\displaystyle
\frac{df}{d\sigma},
\end{eqnarray}
that is $\cal{K}$ works as an effective Hamiltonian for the time $\sigma$.
For $L_z=0$ we have
\begin{eqnarray}
\label{ham2}
\begin{array}{c}
\displaystyle{
{\cal{K}}=\frac{p_u^2+p_v^2}{2}-2
+\frac{(\omega_L n_0^3)^2}{8} u^2 v^2 (u^2+v^2)+} \cr\cr 
\displaystyle{\frac{\epsilon_0}{2}\,(u^4-v^4) 
\sin \phi+\omega_0 \tilde{N} (u^2+v^2).}
\end{array}
\end{eqnarray} 
For $\epsilon=0$, $\omega_0 \tilde{N}=1/2$ and therefore  
the Hamiltonian (\ref{ham2}) represents two 
harmonic oscillators of unit frequency 
coupled by the term $\frac{1}{8}\,(\omega_L n_0^3)^2
u^2 v^2\,(u^2+v^2)$ which originates from the diamagnetic interaction.
For low scaled magnetic field $\omega_L n_0^3$, the motion is regular
and the orbits are quasiperiodic. However, the diamagnetic term
has cylindrical symmetry since it depends only on the perpendicular
distance $\rho$ from the magnetic field axis.
On the contrary, the Coulomb term has spherical symmetry. When these two
terms are of comparable magnitude, 
\begin{eqnarray}
\label{cylsph}
\frac{1}{\sqrt{\rho^2+z^2}}\sim\frac{1}{8}\omega_L^2\rho^2
\end{eqnarray}
(that gives $\omega_L n_0^3$ of the order of unity), then 
the competition between different symmetries leads to chaotic motion
\cite{Delande,Wintgen,Robnik}.
In Fig.\ref{fig1} the classical phase space structure is illustrated 
by the Poincar\'e surfaces of section. 
For $\omega_L n_0^3=3$ some islands of stability still exist
but their size is small, while for $\omega_L n_0^3=9.2$ no
regular structures are visible and the classical motion is dominated by
global chaos \cite{Delande,Wintgen,Robnik}. 

Due to this internal chaos,
the turn on of the microwave field immediately leads to diffusive 
energy excitation of the electron, with classical diffusion rate 
per unit time $D_B=<(\Delta E)^2>/\Delta t$. 
From Eq.(\ref{ham1})
$\dot E=\epsilon z \omega \cos (\omega t)$, with the typical
frequencies for the $z$ motion 
of the order of the Kepler frequency $\omega_K=n_0^{-3}$
(for $\omega_L n_0^3$ not too large with respect to $1$).
For $\omega_0\ll 1$,
$D_B$ can be estimated in quasilinear approximation \cite{mag1},
giving $D_B\sim(\epsilon z \omega)^2/\omega_K$ and then
\begin{eqnarray}
\label{diff1}
D_B \approx \chi_1 D_0 \omega_0^2,
\end{eqnarray} 
where $\chi_1$ is a constant to be numerically determined and
$D_0=\epsilon^2 n_0/2$ is the diffusion rate for $\omega_L=0$ and 
$\omega_0=1$ when the
microwave intensity is strong enough to induce chaos \cite{IEEE}.

For $\omega_0 \gg 1$, in analogy with the case without magnetic field, 
the microwave interaction is mainly effective
when the electron passes close to the nucleus, where the Coulomb 
term dominates the diamagnetic term,
and therefore, as for the case $\omega_L=0$, one has \cite{mag1}:
\begin{eqnarray}
\label{diff2}
D_B\approx \chi_2 D_0 \omega_0^{-4/3},
\end{eqnarray} 
with a constant $\chi_2$ again to be numerically determined. 

\begin{figure}
\centerline{
\epsfxsize=8cm
\epsfysize=8cm
\epsffile{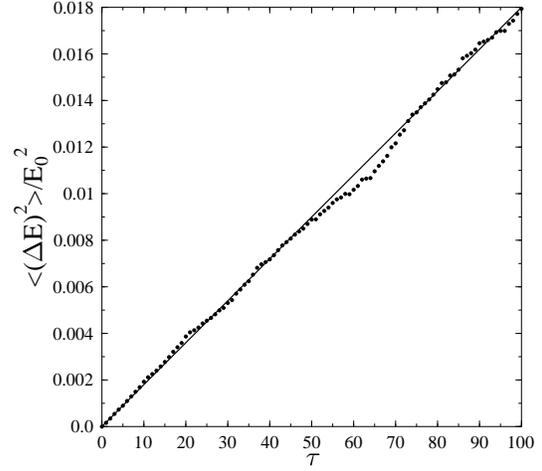}
}
\caption{
Example of diffusive energy excitation $\langle(\Delta E)^2\rangle/E_0^2$
with the number of microwave periods $\tau$, for $\omega_L n_0^3=3$,
$\epsilon_0=0.005$ and $\omega_0=0.1$. The straight line fit allows to
determine $D_B/D_0=0.062$ via the relation $<(\Delta E)^2>=(2\pi D_B/
\omega )\tau$.
}
\label{fig2}
\end{figure}

An example of diffusive energy excitation, for scaled frequency 
$\omega_0\ll 1$, is shown 
in Fig.\ref{fig2}.
Here $\omega_L n_0^3=3$, so that the motion is chaotic even in the
absence of the microwave and therefore a diffusive process occurs even
if the microwave intensity is very small.

\begin{figure}
\centerline{
\epsfxsize=8cm
\epsfysize=8cm
\epsffile{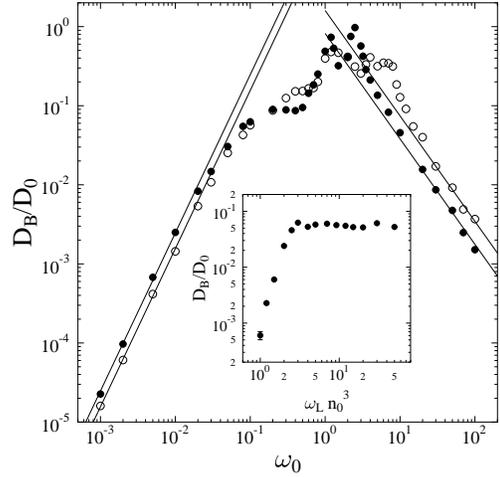}
}
\caption{
Dependence of the scaled diffusion rate $D_B/D_0$ on the scaled 
frequency $\omega_0$ for $\epsilon_0=0.005$, 
$\omega_L n_0^3=3$ (full circles) 
and $\omega_L n_0^3=9.2$ (open circles). The straight lines show the
theoretical dependence for $\omega_0 \ll 1$ (Eq.(\ref{diff1}))
and $\omega_0 \gg 1$ (Eq.(\ref{diff2})),
with $\chi_1=25$, $\chi_2=0.8$ ($\omega_L n_0^3=3$),
$\chi_1=16$, $\chi_2=1.6$ ($\omega_L n_0^3=9.2$). 
Ensembles from $200$ to $1000$ trajectories, initially distributed
microcanonically on the energy shell, have been used.
The insert shows $D_B/D_0$ as a function of the scaled 
Larmor frequency 
$\omega_L n_0^3$, for $\epsilon_0=0.005$ and $\omega_0=0.1$. 
}
\label{fig3}
\end{figure}

The frequency dependence of the diffusion rate $D_B$ is shown in
Fig.\ref{fig3}, for $\omega_L n_0^3=3$ and $9.2$.
The asymptotic behaviors (both for small and large frequencies)
are in agreement with the theoretical estimates (\ref{diff1}) and
(\ref{diff2}) respectively, indicated in the figure by the straight 
lines, with the coefficients $\chi_1$ and $\chi_2$ numerically determined.
These coefficients depend weakly on $\omega_L$, 
for magnetic field strong enough to induce internal chaos
($\omega_L n_0^3 \agt 2$). Below this value the internal motion 
becomes integrable and for low enough $\epsilon_0$ diffusion drops 
to zero, as demonstrated in the insert of Fig.\ref{fig3} (see 
also \cite{mag1}).

The classical diffusive process will lead the electron to ionization
in a time $t_I \approx E_0^2/D_B$, with $E_0=-1/2n_0^2$   the initial 
electron energy. 
In the quantum case, for $n_0\gg 1$
the time evolution initially follows the classical
diffusion. The excitation proceeds via a chain of one--photon
transitions, which eventually brings the electron into
continuum. However, if the ionization time 
is sufficiently large,  quantum interference effects
may suppress the diffusive excitation leading to dynamical 
localization in the number of photons. 

In order to check this possibility we numerically
analyzed the quantum dynamics \cite{mag2}, following the wave packet
evolution in the eigenstates  basis of the magnetic field problem 
with $\epsilon=0$.

To obtain these eigenstates at $\epsilon=0$ in a given energy window around 
the initially excited level with eigenenergy $E_{\lambda_0}\approx E_0$,
we diagonalized the Hamiltonian in a parabolic
Sturmian basis \cite{CCSG87,BD95}. 
This basis is well suited since it is complete and discrete and
can efficiently reproduce both the bound and continuum states 
of the hydrogen atom. 
In addition, all the Hamiltonian matrix elements 
can be expressed in a simple analytical
form and strong selection rules exist for the parabolic quantum numbers
$n_1$, $n_2$ ($\Delta n_i=0,\pm 1,\pm 2$ for $i=1,2$).
A minor disadvantage of dealing with a Sturmian basis is associated
with its nonorthogonality. Due to this fact, we had to solve a
generalized eigenvalue problem of the type $H \psi_i=
E_{\lambda_i} S \psi_i$, where also the overlap matrix $S$ 
has strong selection rules ($\Delta n_i=0, \pm 1$ for $i=1,2$). 
As a result the matrices $H$ and $S$ are very sparse. To study the time
evolution we used up to $1250$ eigenstates of the above generalized 
eigenvalue problem. In order to obtain a good convergence of these
$1250$ eigenvectors and eigenvalues we had to diagonalize matrices
of size larger than $13,000$.
For these reasons the use of an efficient Lanczos algorithm 
\cite{Sorensen} has been crucial. The time evolution was computed by a
split--step metod, similar to the one used in \cite{CCSG87}.

In our computations, we considered as initial state an eigenstate 
at $\epsilon=0$ with eigenenergy 
$E_{\lambda_0}\approx E_0=-1/2n_0^2$ and the time evolution 
in the microwave field was followed up to $\tau=200$
microwave periods. The parameters were varied in the intervals
$0.05\leq \omega_0 \leq 3$, $0.002\leq \epsilon_0 \leq 0.04$,
$40\leq n_0 \leq 80$,
for $\omega_L n_0^3=3$ and $9.2$.

\begin{figure}
\centerline{
\epsfxsize=8cm
\epsfysize=8cm
\epsffile{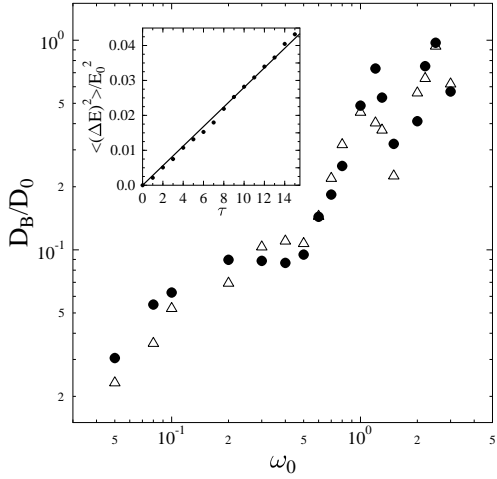}
}
\caption{
Classical (full circles) and quantum (open
triangles) scaled diffusion rates, for $\omega_L n_0^3=3$. 
In the quantum case $n_0=60$,
$0.01 \le \epsilon_0 \le 0.04$.
The insert shows an example of diffusive energy excitation in the quantum
case, for $\omega_L n_0^3=3$, $\epsilon_0=0.02$, $\omega_0=0.1$; the straight 
line fit gives $D_B/D_0=0.055$.
}
\label{fig4}
\end{figure}

In this quasiclassical regime the quantum energy excitation has 
initially a
diffusive character (see the insert of Fig.\ref{fig4}) and
therefore it is possible to compute a quantum diffusion coefficient
$D_B^{(q)}$ from a linear fit, for the few first microwave periods,
of the wave packet energy square variance $<(\Delta E)^2>$
vs. time.
Fig.\ref{fig4} shows that
quantum and classical diffusion coefficients are similar over a
wide range of frequencies ($0.05\leq\omega_0\leq 3$ for $n_0=60$).
This demonstrates that quantum dynamics mimics, for a finite interaction
time, the classical excitation.

\begin{figure}
\centerline{
\epsfxsize=8cm
\epsfysize=8cm
\epsffile{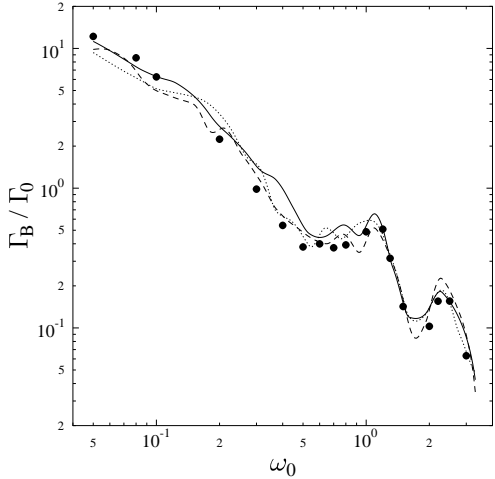}
}
\caption{
Gamma ratio $\Gamma_B/\Gamma_0$ vs. scaled frequency $\omega_0$,
for $\omega_L n_0^3=3$, $n_0=40$ (full line), $n_0=60$ (dotted line),
$n_0=80$ (dashed line). The one--photon transition rate $\Gamma_B$
is evaluated according to Fermi's Golden Rule
(Eq.(\ref{gamma})), with the
Dirac delta function substituted by a Lorentzian function,
with spread $\tilde{\epsilon}=0.5 / \rho_B$, 
where $\rho_B=0.34 n_0^4$ is the density of states. 
Full circles show classical values of $D_B/D_0 \omega_0^2$ at
$\omega_L n_0^3=3$. 
}
\label{fig5}
\end{figure}

In the quasiclassical regime, the one--photon transition rate 
$\Gamma_B$ can be related to the classical diffusion rate \cite{Dima87}
as $\Gamma_B=D_B/\omega^2$. Indeed
the change in energy produced by a one--photon transition is
$\Delta E=\pm\omega$ and $\Gamma_B$ measures the number of such transitions
per unit time.
Therefore, the ratio $\Gamma_B/\Gamma_0$,
where $\Gamma_0=D_0 n_0^6$ is the transition rate for
the chaotic case at $\omega_L=0$ and $\omega_0=1$, is equal to 
\begin{eqnarray}
\label{gammacq}
\frac{\Gamma_B}{\Gamma_0}=\frac{D_B}{D_0\omega_0^2}.
\end{eqnarray}
This result is remarkable as it relates the quantum transition rate to a 
classical characteristic of motion, namely the diffusion coefficient. 
In order to check
the above estimate, $\Gamma_B$ was numerically evaluated according
to Fermi's Golden Rule:
\begin{eqnarray}
\label{gamma}
\begin{array}{c}
\displaystyle{
\Gamma_B=\frac{\pi}{2}\,\epsilon^2
\sum_{m \neq n} |z_{mn}|^2 \{\delta(E_{\lambda_m}-
E_{\lambda_n}+\omega)+}\cr\cr
\displaystyle{\delta(E_{\lambda_m}-E_{\lambda_n}-\omega)\},}
\end{array}
\end{eqnarray} 
with $z_{mn}$ matrix elements of $z$ between the eigenstates at $\epsilon=0$
with eigenenergies $E_{\lambda_m},E_{\lambda_n}$.
For the numerical computation, 
the Dirac delta function in Eq.(\ref{gamma}) was substituted by a 
Lorentzian function:
$\delta(x) \to f(x)=(1/\pi)(\tilde{\epsilon}\,/(x^2+\tilde{\epsilon}^2))$, 
with $\tilde{\epsilon}$ of the order of the level spacing.
The validity of the classical--quantum correspondence 
(\ref{gammacq}) is confirmed in Fig.\ref{fig5},
for $n_0=40,60,80$ and $0.05\leq\omega_0\leq 3$.  

Even if initially the interaction with the microwave field results in
a quantum diffusive excitation, 
quantum interference effects can suppress the diffusive behavior 
before ionization takes place.
In this case, the diffusive broadening of the
quantum probability distribution over unperturbed levels stops.
The corresponding localization length (measured in number of absorbed photons)
$\ell_B$ is proportional to the one--photon transition rate $\Gamma_B$
and to the density of coupled states $\rho_B$: $\ell_B=2\pi \Gamma_B
\rho_B$ \cite{Dima87}. For the chaotic case at $\omega_L=0$ and 
$\omega_0=1$ the
localization length is $\ell_\phi=3.3\epsilon_0^2n_0^2=2\pi D_0 n_0^6 
\rho_0=2\pi\Gamma_0\rho_0$, where $\rho_0=n_0^3$ is the density of 
coupled states \cite{IEEE}. 
Actually, without the magnetic field there is an
additional approximate integral of motion,
related to Coulomb degeneracy, and therefore 
the density of coupled states $\rho_0$ is by a factor $n_0$ smaller 
than the number of levels 
per unit energy interval $n_0^4$ \cite{IEEE}. 
Therefore from the above expressions and Eq.(\ref{gammacq}) 
we get \cite{mag1,mag2}
\begin{eqnarray}
\label{locb}
\ell_B=\ell_\phi\frac{D_B}{D_0\omega_0^2}\frac{\rho_B}{\rho_0}.
\end{eqnarray} 
This result provides a theoretical formula for the 
the quatum localization length, which involves 
only classical characteristics of motion, namely
the classical diffusion coefficient and the density of coupled 
states. The latter is related, in the quasiclassical regime, to
the phase space structure (see below
Eqs.(\ref{volume})--(\ref{density})).
The estimate (\ref{locb}) is valid for $\ell_B>1$, 
namely in the quasiclassical 
regime, in which a large number of photons is absorbed and a large number
of levels is excited. Moreover, the microwave frequency should be larger
than the average level spacing ($\omega \rho_B >1$); otherwise, levels
would move adiabatically in time leaving no room for diffusive
energy growth \cite{IEEE}.

\begin{figure}
\centerline{
\epsfxsize=8cm
\epsfysize=8cm
\epsffile{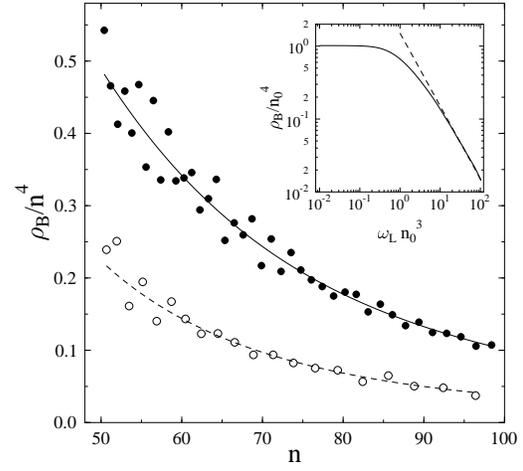}
}
\caption{
Scaled density of states $\rho_B/n^4$ as a function of $n$ 
for $\omega_L n_0^3=3$
(full circles) and $\omega_L n_0^3=9.2$ (open circles) at $n_0=60$.
Solid and dashed lines refer to the quasiclassical computation
(Eq.(\ref{density})). The insert gives
quasiclassical scaled density of states vs. scaled Larmor
frequency. The straight line shows the theoretical  asymptotical 
dependence ($\rho_B/n_0^4=c/\omega_L n_0^3$, with $c=1.5$).
}
\label{fig6}
\end{figure}

For $\omega_L n_0^3$ sufficiently large, the internal motion is chaotic:
as a consequence, Coulomb degeneracy is removed 
and the density of coupled states $\rho_B\sim n_0^4$. More precisely,
$\rho_B=0.34 n_0^4$ (for $\omega_L n_0^3=3$, $n_0=60$) and  
$\rho_B=0.14 n_0^4$ (for $\omega_L n_0^3=9.2$, $n_0=60$)
(see Fig.\ref{fig6}). 
The density follows the dependence 
$\rho_B\propto 1/\omega_L$. This is related to the fact that the
diamagnetic term is identical to a two--dimensional harmonic
oscillator in the $x-y$ plane. Indeed 
for a harmonic oscillator the spacing between levels
is proportional
to the frequency (here represented by $\omega_L$) and therefore
their density scales with $\omega_L^{-1}$.
This result is confirmed by a quasiclassical computation of the 
density of states (see the insert of Fig.\ref{fig6}).
Here the number $\Omega(E)$ of quantum states having energy less
than $E$ is related to the corresponding volume in phase space
\begin{eqnarray}
\label{volume}
\displaystyle{
\Omega(E)=\frac{1}{(2\pi)^2}\int_{H<E}\,d\rho dz dp_\rho dp_z,
}
\end{eqnarray}
where the Hamiltonian $H$ is taken at $\epsilon=0$.
The density of states is 
\begin{eqnarray}
\label{density}
\rho_B=\frac{d\Omega(E)}{dE}=n_0^4\,
\frac{d\tilde{\Omega}(\tilde{E})}{d\tilde{E}},
\end{eqnarray}
where the volume in the scaled phase space is defined as
\begin{eqnarray}
\label{volscal}
\displaystyle{
\tilde{\Omega}(\tilde{E})=\frac{1}{(2\pi)^2}
\int_{\tilde{H}<\tilde{E}=-\frac{1}{2}}\, du dv dp_u dp_v,
}
\end{eqnarray}
with
\begin{eqnarray}
\label{hscal}
\displaystyle{
\tilde{H}=\frac{p_u^2+p_v^2}{2(u^2+v^2)}-\frac{2}{u^2+v^2}
+\frac{(\omega_L n_0^3)^2}{8}\,u^2 v^2.
}
\end{eqnarray}
According to the results of Fig.\ref{fig6} such semiclassically determined
density of states is in a good agreement with the results obtained by
direct diagonalization of hydrogen in magnetic field.

According to the estimate (\ref{locb}) and differently from the zero 
magnetic field case, the localization length $\ell_B$ is non
homogeneous in the number $N_\phi$ of absorbed photons.
Indeed, from equations (\ref{diff1}) and (\ref{diff2}) we get
$\ell_B\sim n_0^{11} \sim (N_I-N_\phi)^{-11/2}$ for $\omega_0\ll 1$
and $\ell_B\sim n_0 \sim (N_I-N_\phi)^{-1/2}$ for $\omega_0\gg 1$
\cite{mag1}, with $N_I=n_0/2\omega_0$ the number of photons required for
ionization and $N_\phi = N - N_0$, where $N_0=E_0/\omega$.
However, for $\ell_B\ll N_I$ this effect is not important and the
quasistationary distribution is exponentially localized in the
photon number $N_\phi$. As for the case of hydrogen in a microwave
field only \cite{IEEE}, the probability distribution over unperturbed
levels ($\epsilon=0$) displays a chain of equidistant peaks in 
energy. The probability amplitudes in a one--photon interval decay 
esponentially with the photon number: 
$\psi_N\propto\exp(-|N-N_0|/\ell_B)$.

The value of the localization length $\ell_B$ is strongly enhanced
compared to the length $\ell_{\phi_\omega}=3.3\epsilon_0^2 n_0^2/
\omega_0^{10/3}$ \cite{IEEE} at $\omega_L=0$ and $\omega_0>1$. 
Actually, for a strong enough scaled magnetic field, Coulomb 
degeneracy is removed and the eigenfunctions $\psi$, when developed
on the basis of the hydrogenic eigenfunctions $\varphi$,
have a large number of randomly 
fluctuating components, increased by a factor $M=\rho_B/\rho_0$:
\begin{eqnarray}
\psi_i\approx\sum_{k=1}^M c_{ik} \varphi_k,
\end{eqnarray}
with $c_{ik}\sim 1/\sqrt{M}$
due to normalization condition.
Because of this, the external microwave perturbation $V$ couples
the eigenstates with typical interaction matrix elements 
\begin{eqnarray}
V_{int}=\langle\psi_j|V|\psi_i\rangle\approx
\sum_{k,l=1}^M c_{jl}^* c_{ik} \langle\varphi_l|V|\varphi_k\rangle.
\end{eqnarray}
Due to selection rules for $V$ in the hydrogenic basis, $V_{int}$
is the sum of order $M$ uncorrelated terms  
and is therefore reduced by a factor $1/\sqrt{M}$ in comparison with 
the case without
magnetic field,
whereas the spacing $\Delta E$
between nearest neighbors levels in the spectrum scales as
$1/M$. Thus, the admixture factor 
\begin{equation}
\eta \sim V_{int}/\Delta E \sim \sqrt{M} \sim \sqrt{n_0}
\end{equation}
is strongly enhanced, for high $n_0$ values, in comparison with
the zero magnetic field case.   
As a result, the localization
length $\ell_B$, which is proportional to $\eta^2$,
is increased \cite{mag2} by a factor 
\begin{equation}
\ell_B/\ell_{\phi_\omega}
\sim \chi_2\rho_B/\rho_0 \sim M \sim n_0.
\end{equation}  
A similar effect was discussed for two interacting particles in
a disordered solid state system \cite{Dima94}: the interaction allows
the two particles to propagate coherently on a distance much larger
than one--particle localization length.
The statistical enhancement effect is quite general and takes place
in different pysical problems: actually this effect was first studied
for the weak interaction parity--breaking in the scattering of 
polarized neutrons on
complex nuclei \cite{Sushkov-Flambaum}. In the compound nucleus the
level spacing is typically $\Delta E\approx 1$ eV, whereas the
typical energy scale due to the strong interaction is 
$\Delta E_0\approx 1$ MeV.
As a consequence, the number of principal components of the
eigenfunctions in the one-particle basis is $M\approx 10^6$ and
therefore the admixture factor between states of opposite parity
is enhanced by a factor $\sqrt{M}\approx 10^3$ compared to a usual
parity breaking admixture of the order $10^{-7}$.

The condition $\ell_B=N_I$ allows to determine the quantum delocalization
border \cite{mag1,mag2}:  
\begin{eqnarray}
\label{deloc}
\epsilon_q=\frac{1}{n_0}\sqrt{\frac{D_0 \omega_0^2}{6.6 D_B}
\,\frac{\rho_0 n_0}{\rho_B \omega_0}}.
\end{eqnarray}
For $\epsilon_0>\epsilon_q$ localization effects
become unimportant and diffusive ionization close to the classical case
takes place.
For $\omega_0\gg 1$ this border is $(\chi_2 n_0/\omega_L n_0^3)^{1/2}$
times smaller than the quantum delocalization border at zero 
magnetic field \cite{IEEE}
$\epsilon_{q0}=\omega_0^{7/6}/\sqrt{6.6 n_0}$. For $\omega_0\ll 1$
the border is given by $\epsilon_q=(\omega_L n_0^3/6.6\chi_1 \omega_0)^{1/2}/
n_0$, which remains well below the ionization border for a static electric
field in the presence of a parallel magnetic field.
Since the localization length is non homogeneous in the number
of absorbed photons and becomes larger near the ionization border, the
actual value of the delocalization border $\epsilon_q$ will be slightly 
decreased as compared to (\ref{deloc}). 

\begin{figure}
\centerline{
\epsfxsize=10cm
\epsfysize=15cm
\epsffile{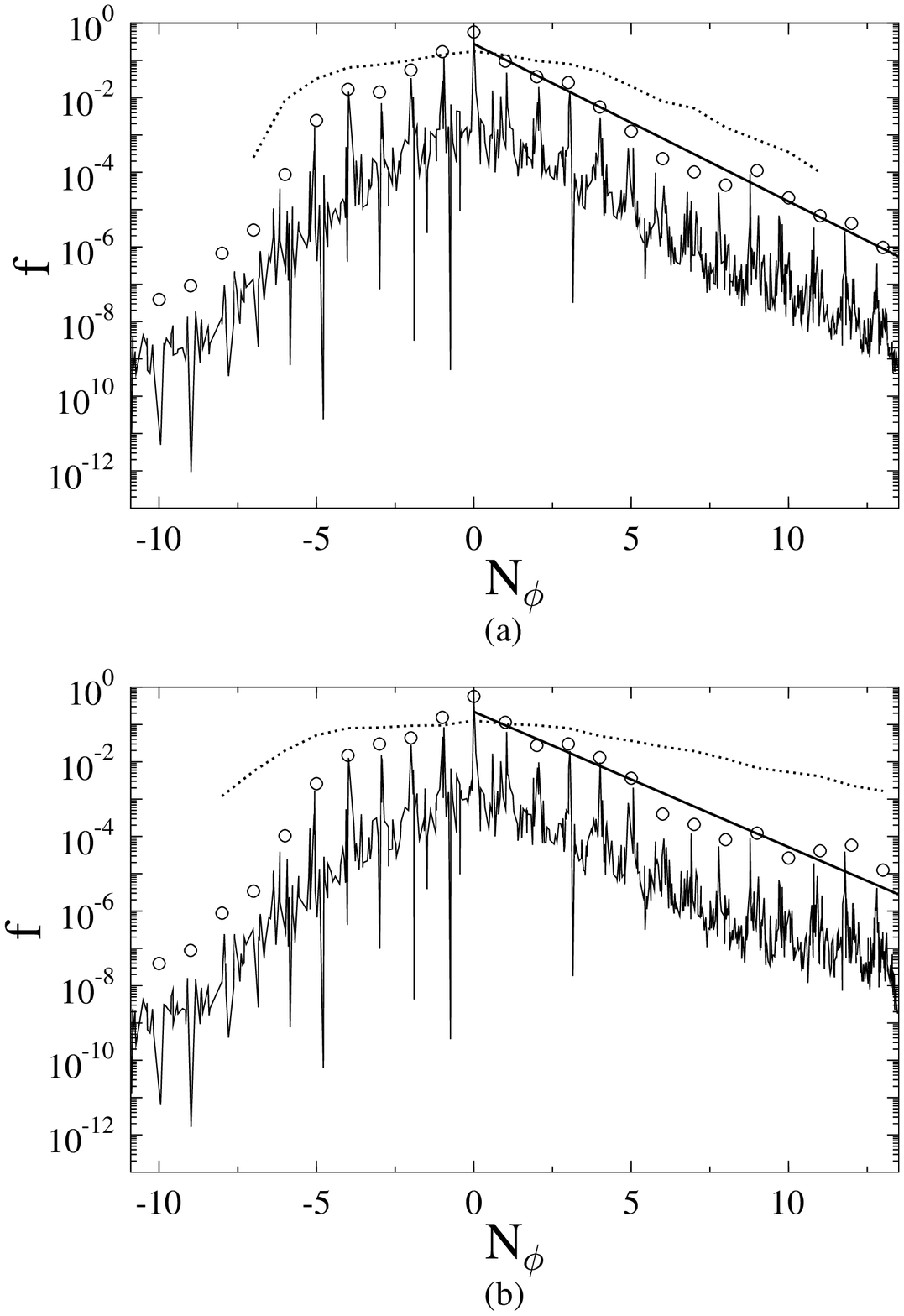}
}
\caption{
Probability distribution vs. number of absorbed photons $N_\phi$:
quantum distribution $f_\lambda$ over the eigenbasis at $\epsilon=0$
(full curve); quantum probability in one--photon intervals $f_N$ (circles);
classical probability in one--photon intervals (dotted line). The
straight line shows the fit for the exponential localization. 
Parameter values: $n_0=50$, $\omega_0=1.3$, $\omega_L n_0^3=3$,
$\epsilon_0=0.008$; scaled diffusion rate $D_B/D_0=0.53$, 
theoretical localization length estimate (Eq.(\ref{locb})) $\ell_B=2.8$;
(a) $80\leq\tau\leq 100$, $\ell_{BN}=2.1$; (b) $180\leq\tau\leq 200$, 
$\ell_{BN}=2.4$.
In the classical simulation $1000$ trajectories are included in the
microcanonical ensemble.
}
\label{fig7}
\end{figure}

\begin{figure}
\centerline{
\epsfxsize=10cm
\epsfysize=15cm
\epsffile{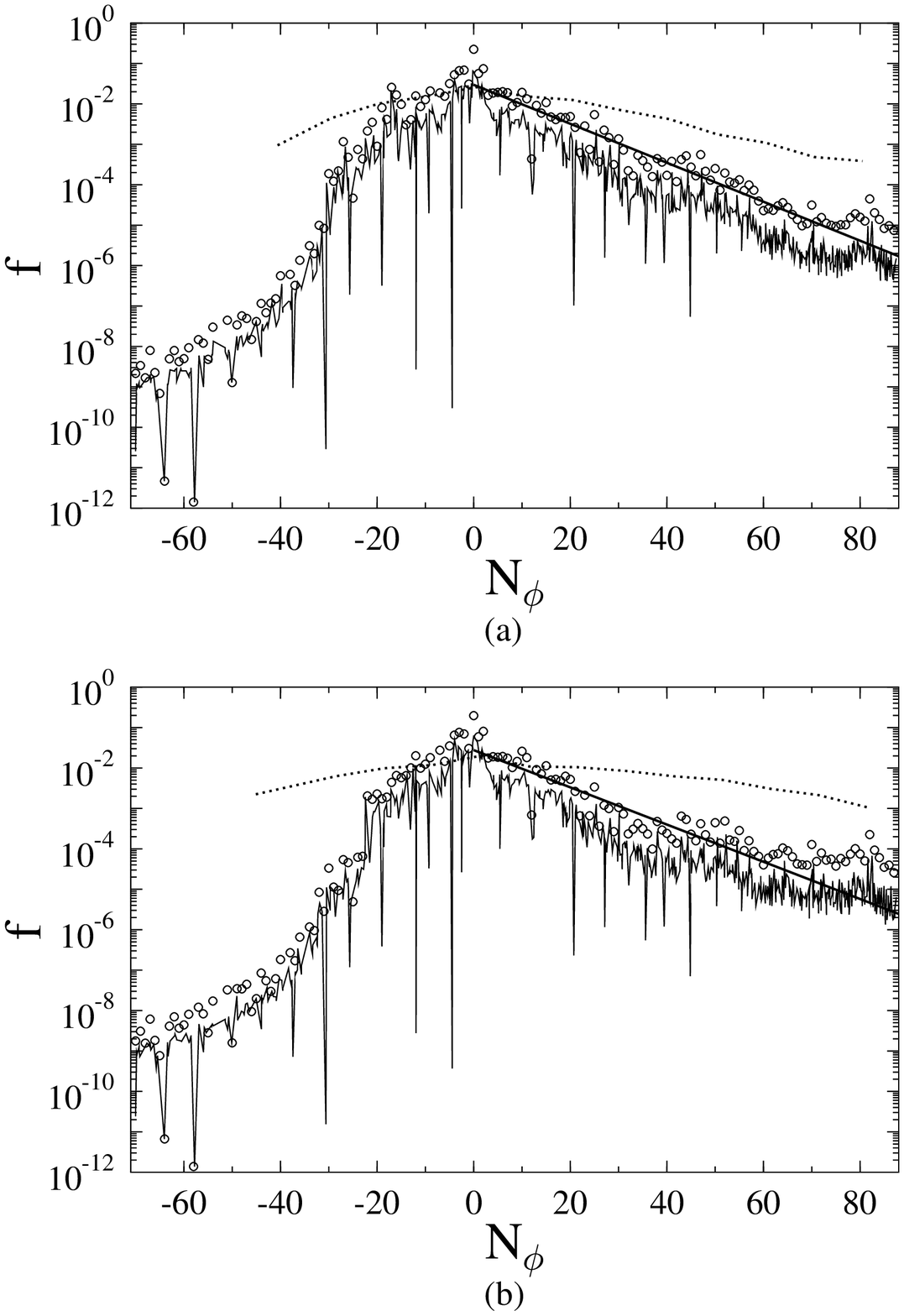}
}
\caption{
Same as in Fig.7, with  
$n_0=50$, $\omega_0=0.2$, $\omega_L n_0^3=3$,
$\epsilon_0=0.007$, $D_B/D_0=0.090$, $\ell_B=15.4$,
(a) $80\leq\tau\leq 100$, $\ell_{BN}=18$; (b) $180\leq \tau\leq 200$, 
$\ell_{BN}=18.8$
(in both cases the fit is limited to $0\leq N_\phi \leq 60$).
The classical ensemble consists of $1000$ trajectories. 
}
\label{fig8}
\end{figure}

The above dynamical localization theory was tested by detailed numerical
simulations of quantum evolution.
The probability distribution $f_\lambda$ over the eigenstates 
at $\epsilon=0$ for the localized case        
is shown in Figs.\ref{fig7},\ref{fig8} as a 
function of the number of absorbed photons $N_\phi=(E_\lambda-
E_0)/\omega$. In order to suppress fluctuactions this probability
was averaged over $20$ microwave periods.
To find the numerical value $\ell_{BN}$ of the localization length
we first computed the probability $f_N$ in each one--photon
interval around integer values of $N_\phi$.
Then the least square fit with $f_N\sim \exp(-2 N_\phi/\ell_{BN})$
for $N_\phi\geq 0$ allows to determine $\ell_{BN}$.
The quantum distribution at $\omega_0=0.2$ (Fig.\ref{fig8}(b))
starts to deviate from the exponential profile for $N_\phi>60$ and a 
plateau appears. This can be related  
to the fact that localization length is non homogeneous 
in the number of absorbed photons ($\ell_B\sim (N_I-N_\phi)^{-11/2}$
for $\omega_0\ll 1$).
In the figures also the classical distribution, normalized to
one--photon interval, is shown. This distribution was obtained
by numerical solution of Hamilton equations with $1000$ trajectories
initially distributed microcanonically on the energy surface at
energy $E_0$.
The comparison of the probability distributions at different times
($80\leq\tau\leq 100$ in Figs.\ref{fig7}(a),\ref{fig8}(a) and
$180\leq\tau\leq 200$ in Figs.\ref{fig7}(b),\ref{fig8}(b)) shows 
that the quantum case is well localized over a quasistationary 
distribution. Its profile does not change significantly by
increasing the interaction time, apart from a slight increase in
the tail.
We attribute the decrease in the probability decay 
at large $N_\phi > 0$ to the growth of
the diffusion rate $D_B$ when $\omega_0$ approaches $1$ from
below (see Fig.\ref{fig3}).

\begin{figure}
\centerline{
\epsfxsize=8cm
\epsfysize=8cm
\epsffile{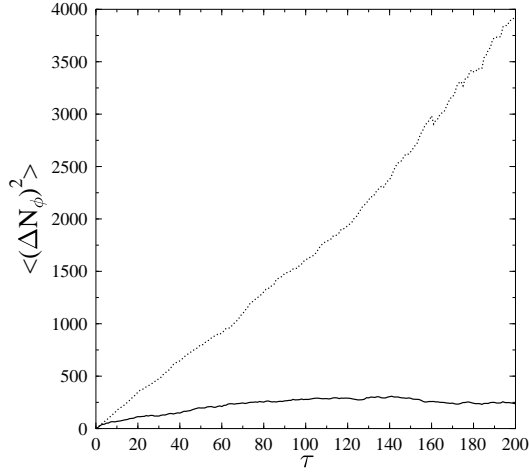}
}
\caption{
Dependence of the quantum (full curve) and classical (dotted curve) 
photon number square variance $<(\Delta N_\phi)^2>$ on the
number of microwave periods $\tau$ for $n_0=60$, $\omega_0=0.1$,
$\epsilon_0=0.005$, $\omega_L n_0^3=3$. The classical ensemble consists of
$5000$ trajectories.
}
\label{fig9}
\end{figure}

These distributions, together with the time dependence of 
the square variance of the
photon number $<(\Delta N_\phi)^2>$ (see Fig.\ref{fig9}) demonstrate
how the quantum--mechanical energy diffusion is strongly suppressed
in comparison with the classical case. 

\begin{figure}
\centerline{
\epsfxsize=8cm
\epsfysize=8cm
\epsffile{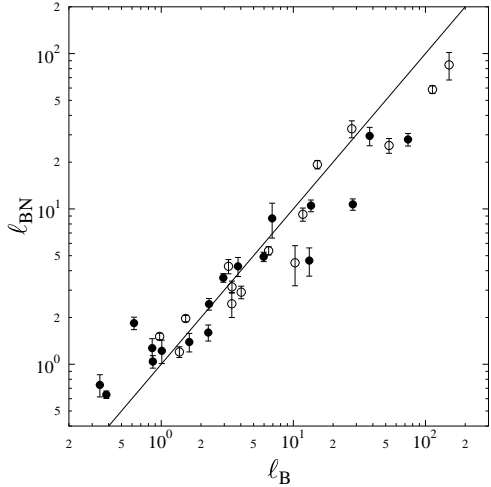}
}
\caption{
The numerically computed localization length $\ell_{BN}$ vs. the 
theoretical estimate $\ell_B$ (Eq.(\ref{locb})) for $\epsilon_0=0.005$ 
(full circles)
and $\epsilon_0=0.01$ (open circles). Data are for $\omega_L n_0^3=3$
or $9.2$, $n_0=60$, $0.05\leq\omega_0\leq 3$.
The straight line corresponds to $\ell_{BN}=\ell_B$.
}
\label{fig10}
\end{figure}

The numerically obtained localization lengths
$\ell_{BN}$ are presented
Fig.\ref{fig10}, for two fixed values of $\epsilon_0$ and different
$\omega_0$ in the range $0.05\leq\omega_0\leq 3$. 
Since $n_0=60$, the number of photons required for ionization goes 
from $N_I=10$ to $N_I=600$.
The straight line,
corresponding to $\ell_{BN}=\ell_B$, demonstrates the fairly good 
agreement between numerical data and dynamical localization theory
over a wide range of parameters (the localization 
length changes by two orders of magnitude).

\begin{figure}
\centerline{
\epsfxsize=8cm
\epsfysize=8cm
\epsffile{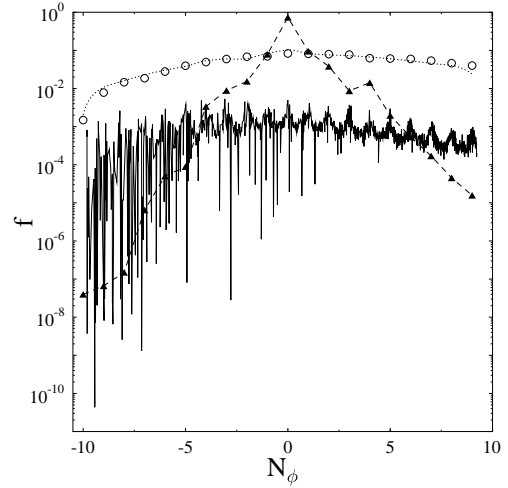}
}
\caption{
Chaotic enhancement of microwave excitation. Same as in Fig.\ref{fig7}, with
$n_0=66$, $\omega_0=2.5$, $\omega_L n_0^3=3$,
$\epsilon_0=0.04>\epsilon_q=0.016$, $D_B/D_0=0.97$, $40\leq\tau\leq 50$. 
Full triangles, linked by a dashed line, give quantum probability 
in one--photon intervals at the same parameter values but
without magnetic field, $290 \leq \tau \leq 300$
and with initial value of the orbital momentum
quantum number $l_0=5$.
In the latter case, 
$\epsilon_c=0.015<\epsilon_0<\epsilon_{q0}=0.14$,
theoretical 
$\ell_{\phi_\omega}=1.1$, numerical $\ell_{\phi_\omega}=1.7$.
}
\label{fig11}
\end{figure}

\begin{figure}
\centerline{
\epsfxsize=8cm
\epsfysize=8cm
\epsffile{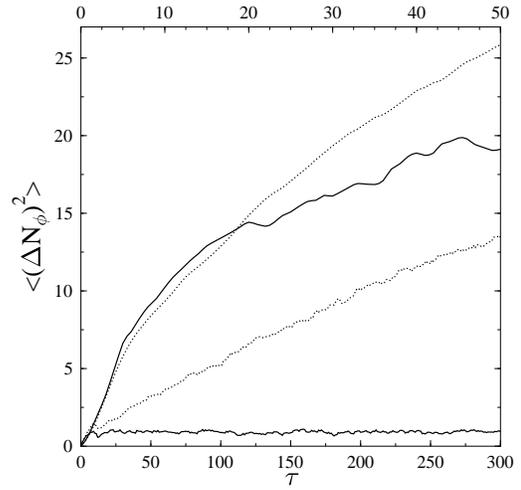}
}
\caption{
Quantum (full curve) and classical (dotted curve) 
square variance of the photon number vs.
the number of microwave periods for the case of Fig.\ref{fig11}. 
Above and upper scale: $\omega_L n_0^3=3$, 
below and lower scale: $\omega_L n_0^3=0$.
The case $\omega_L=0$ is followed for a longer interaction time
than for $\omega_L n_0^3=3$ to emphasize dynamical localization effects. 
}
\label{fig12}
\end{figure}

Finally, Figs.\ref{fig11},\ref{fig12}
show that the excitation at $\omega_L n_0^3=3$ is much stronger
than at zero magnetic field, due to the chaotic enhancement
of electron's interaction with the microwave field.
Since the condition for the transition to chaotic dynamics
at zero magnetic field is given by
\cite{CCSG87,IEEE} $\epsilon_0>\epsilon_c=1/49\omega_0^{1/3}$,
the case of Fig.\ref{fig11} (that corresponds to Fig.10 in \cite{IEEE})
is above the 
chaos threshold ($\epsilon_0>\epsilon_c=0.015$). However the quantum 
distribution at $\omega_L=0$ clearly demonstrates exponential 
localization of diffusive excitation since $\epsilon_0<\epsilon_{q0}=0.14$.
On the contrary, for $\omega_L n_0^3=3$ we are above the delocalization
border ($\epsilon_0>\epsilon_q=0.016$)
and thus numerical results show a good agreement
between classical and quantum distributions. 
The quantum and classical square variance of the photon number are
shown in Fig.\ref{fig12}. In the absence of the magnetic field, the
localization of the quantum motion after a few microwave periods is 
evident; on the contrary, for $\omega_L n_0^3=3$ the quantum square
variance follows the classical one and starts to deviate only
for $\tau>20$ because of the finite size of the basis. 

\section{Rydberg atoms in static electric and microwave fields}
Another way to study the microwave interaction in an 
intrinsically chaotic atomic system is to consider alkali Rydberg atoms
in a static electric field; this case is interesting as it is more
suitable for laboratory experiments.

The Hamiltonian of a hydrogen atom in a static electric field is separable 
in parabolic coordinates and thus the classical motion is integrable. 
On the contrary, in alkali atoms the valence electron moves in an 
attractive potential in which the $1/r$ dependence is modified
near the origin due to the charge distribution of the inner electrons,
which form a core of size a few Bohr radius.
Due to precession in a static electric field, the electron always
collides with the core. 
This fact prevents the separation of variables and the motion
becomes chaotic for certain field and energy values
\cite{Kleppner}.

The potential within the many--electron core region can be reasonably
approximated by a one--electron spherically symmetric model
\cite{Dando}:
\begin{eqnarray}
\label{corpot}
V_c(r)=-\frac{Z-1}{r}\exp(-\alpha_1 r)-\alpha_2\exp(-\alpha_3 r),
\end{eqnarray}
where $Z$ is the nuclear charge and the parameters $\alpha_i$ 
$(i=1,2,3)$ are
numerically determined so as to reproduce quantum defects, that 
characterize the ionic core of alkali atoms in quantum mechanics.
In this way, the overall potential $V(r)=-1/r+V_c(r)$ complies with
the two conditions $V(r)\sim -Z/r$ (for $r\to 0$, where the
core fails to screen the nuclear charge) and 
$V(r)\sim -1/r$ (for $r\to\infty$, where due to the screening
the valence electron feels only a unit charge).
Even if different core potential forms could reproduce quantum
defects, both classical and quantum properties of the system are
sensitive only to the size of these defects, not to details of the 
core.

In the quantum case, the core potential $V_c(r)$ introduces a
phase shift $\pi\delta_l$ in the radial wave function. The
electron energy is then given by $E_{nl}=-1/2 (n-\delta_l)^2$, namely
the effective principal quantum number becomes $n^{\star}=n-\delta_l$,
where the quantum defect $\delta_l$ depends on the orbital momentum
$l$ but changes only weakly with the energy $E$. 
Since the size of the core is small only the quantum defects for
low angular momentum values ($l<3$) are significantly different from
zero.

At the ionization threshold, the quantum defects $\delta_l$ are
given, in quasiclassical approximation, by the action difference 
in the nonhydrogenic and hydrogenic radial motion 
between turning points \cite{Wunner}:
\begin{eqnarray}
\label{quantumdef}
\begin{array}{c}
\displaystyle{
\delta_l=\frac{\sqrt{2}}{\pi} \lim_{R\to\infty}
\left[\int_{r_{0c}}^R\,\sqrt{\frac{1}{r}-V_c(r)-
\frac{(l+1/2)^2}{2 r^2}} \,dr\right.}\cr\cr
\displaystyle{-\left.\int_{r_0}^R\,\sqrt{\frac{1}{r}-
\frac{(l+1/2)^2}{2 r^2}} \,dr\,\right].}
\end{array}
\end{eqnarray}
In this formula, $l(l+1)$ has been replaced by $(l+1/2)^2$ so that 
quasiclassical approximation remains valid also for small angular
quantum numbers \cite{Landau}.
The turning point for the hydrogenic part is $r_0=(l+1/2)^2/2$,
while for the nonhydrogenic part we obtain $r_{0c}$ as the root of
$V(r_{0c})+(l+1/2)^2/2r_{0c}^2=0$. 
The parameters $\alpha_i$ were fixed in order to
reproduce the most important quantum defects 
$\delta_0,\delta_1,\delta_2$. For Li we assumed $\delta_0=0.4$,
$\delta_1=0.04$, $\delta_2=0$; for Na $\delta_0=1.35$,
$\delta_1=0.85$, $\delta_2=0.01$; for Rb $\delta_0=3.14$,
$\delta_1=2.65$, $\delta_2=1.35$; we always considered 
$\delta_{l>2}=0$ \cite{qdeftab}. Typical values of 
the $\alpha_i$ parameters are quite different from those
given in Tab.1 of \cite{Dando}, probably because in that case  
these parameters were optimized so as to directly reproduce the 
energy levels. 
In any case this difference is not important in evaluating 
the quantum defects, that can be reproduced from 
Eq.(\ref{quantumdef}) with $\alpha_i$ parameters taken from
\cite{Dando} with a negligible error. 

The classical Hamiltonian for alkali atoms in parallel static electric 
field $\epsilon_s$ and microwave field $\epsilon\sin(\omega t)$,
both in the $z$--direction, writes as
\begin{eqnarray}
\label{hamqdef}
H=\frac{p^2}{2}-\frac{1}{r}+V_c(r)+\epsilon_s z+\epsilon z
\sin(\omega t).
\end{eqnarray}

The scaling property expressed by Eq.(\ref{scaled}) remains valid,
provided that also the static field and the core potential parameters
are scaled as $\epsilon_{s0}\equiv\epsilon_s n_0^4$,
$\alpha_{i0}\equiv\alpha_i n_0^2$ ($i=1,\ldots,3$).
In other words, differently from the case of hydrogen,
the problem depends separately on $\epsilon_s$
and $n_0$ and not only on the scaled parameter $\epsilon_{s0}$.
As a matter of fact, the 
transformation $\alpha_i\to\alpha_{i0}=\alpha_i n_0^2$
changes the rescaled core size. 

The study of the classical dynamics of the above problem
was done in a similar way to the case of the hydrogen atom in magnetic
and microwave fields. The system was studied in an
extended phase space that includes the time as a new canonical
variable and the Coulomb singularity at $r=0$ 
was removed by the introduction
of the semi--parabolic coordinates $u,v$ and the regularized time
$\sigma$ (Eq.(\ref{scaltau})). 
For simplicity's sake, we considered only the case with 
$z$--component of the angular momentum $L_z=0$.

\begin{figure}
\centerline{
\epsfxsize=8cm
\epsfysize=8cm
\epsffile{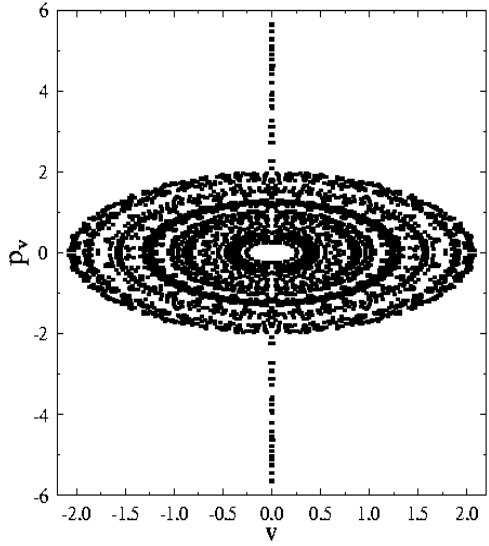}
}
\caption{
Poincar\'e surface of section for Na at 
$\epsilon_{s0}=0.02$,
$n_0=60$. The $v-p_v$ plane of section is defined by $u=0$, $p_u>0$.
The section is constructed following only one trajectory.
Core potential parameters are 
$\alpha_1=2.48$, $\alpha_2=0.54$,
$\alpha_3=1.43$.
}
\label{fig13}
\end{figure}

For $\epsilon=0$, due to the core effects,
we have a transition from regular to chaotic motion by increasing
the static field $\epsilon_{s0}$ \cite{Kleppner}.
However, the nature of chaotic dynamics in this case is qualitatively
different from the case of diamagnetic 
hydrogen. In the latter case, the irregular motion is due to the 
competition of the spherical Coulomb interaction, that dominates near 
the nucleus, and the cylindrical diamagnetic interaction, that dominates 
at large distances. In the former case instead, the chaotic motion 
arises from the presence of a 
core, which is small compared to the typical size of 
Rydberg atoms. Fig.\ref{fig13} 
shows that the phase space remains dominated by orbits which
jump between different tori of the Coulomb problem. A trajectory 
follows a torus of the hydrogenic system until it encounters the core 
region where it is scattered on a different torus. 
As a result a single trajectory is able to explore most of the phase 
space energetically accessible to it. 
The lobes near $v=0$ in the Poincar\'e surface of section 
in Fig.\ref{fig13} are due to the attractive
core, which allows larger values of the momentum for a given energy. 

This chaotic motion affects also the quantum energy level spacing statistics
\cite{Bohigas}. For integrable systems the level spacing statistics
has generally the Poisson form
$P(s)=\exp(-s)$, with $s$ the spacing between two consecutive 
levels, normalized to the local mean level spacing. For systems
that display chaotic classical motion the distribution $P(s)$ 
is described by the Random Matrix Theory (RMT).
For a number of chaotic systems, with time reversal symmetry, the
energy level statistics has been found in agreement with the 
predictions of the Gaussian orthogonal ensemble of random matrices,
with $P(s)$ given by the Wigner--Dyson distribution
\begin{eqnarray}
\label{Wigner}
P(s)=\frac{\pi}{2}\,s\exp(-\frac{\pi}{4}s^2).
\end{eqnarray}

\begin{figure}
\centerline{
\epsfxsize=8cm
\epsfysize=8cm
\epsffile{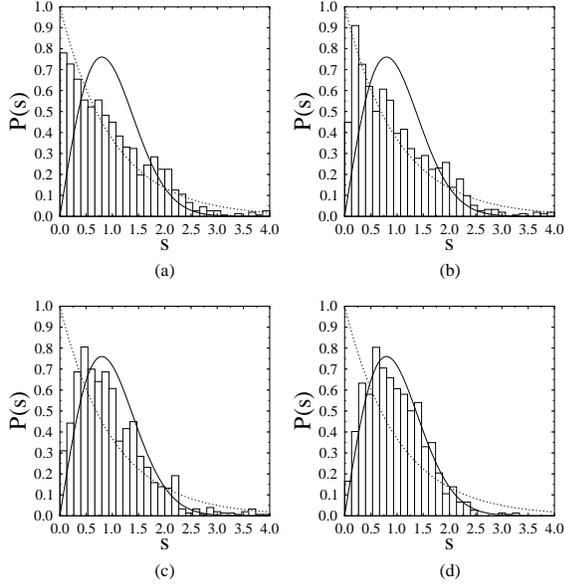}
}
\vspace{0.5cm}
\caption{
Level spacing statistics $P(s)$ for alkali atoms in a static electric
field, with $\epsilon_{s0}=0.02$, $n_0=60$, 
$55 \leq n \leq 72$, for (a) H, (b) Li,
(c) Na and (d) Rb. The dotted line shows the Poisson distribution, 
the solid line is the Wigner--Dyson distribution (Eq.(\ref{Wigner})).
}
\label{fig14}
\end{figure}

A Wigner-Dyson distribution of level spacings 
has been seen in the  Li atom in a static 
electric field strong enough to induce classical chaos \cite{Kleppner}. 
In Fig.\ref{fig14} the normalized nearest 
neighbor distribution $P(s)$ is shown for $\epsilon_{s0}=0.02$ at 
$n_0=60$, in the range $55 \leq n \leq 72$. 
The hydrogen displays nearly Poissonian behavior while for Rb,
in which quantum defects are large, $P(s)$ is close to the
RMT results. The case of Na demonstrates
a weaker level repulsion than Rb, while Li, which has smaller 
quantum defects, presents a distribution rather close to the Poisson
case.

The energy levels for alkali atoms in a static electric field
(the so--called Stark levels)
were obtained by matrix diagonalization in a spherical hydrogenic
basis; quantum defects were introduced simply in the alkali energy levels 
(for angular momentum $l<3$), while dipole matrix elements were taken as
in the hydrogen case. This approach is valid outside
the ionic core, where $V(r)$ is essentially hydrogenic and so is
particularly well suited for Rydberg states, in which the electron
mainly remains far from the 
nucleus. The off--diagonal dipole matrix elements for the $z$ operator
decrease rapidly with the difference in energy between initial and final
states. As a result, Stark eigenvalues and eigenvectors in a given
energy window can be obtained by diagonalization of Hamiltonian
matrices of limited size 
around this region. A limitation of the method is that only bound 
states are included in the basis set and increasingly large matrices 
must be used to obtain convergence as the static field ionization
border $\epsilon_s n^4=0.13$ is approached.

\begin{figure}
\centerline{
\epsfxsize=8cm
\epsfysize=8cm
\epsffile{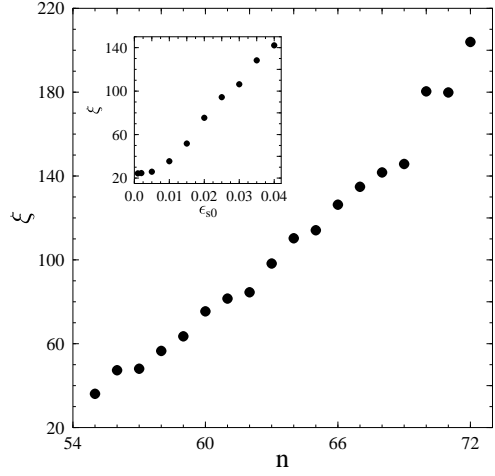}
}
\caption{
Inverse participation ratio $\xi$ (Eq.(\ref{IPR})) for Rb, at 
$\epsilon_{s0}=0.02$ and $n_0=60$, as a function of $n$. The insert shows 
the dependence of $\xi$ on $\epsilon_{s0}$, for $n_0=60$.
}
\label{fig15}
\end{figure}

The number of mixed hydrogenic states can be characterized by the
inverse participation ratio
\begin{eqnarray}
\label{IPR}
\xi_{\lambda_i}=(\sum_{n,l} |c_{\lambda_i}(n,l)|^4)^{-1},
\end{eqnarray}
where $c_{\lambda_i}(n,l)$ are the coefficients of the expansion of the
Stark eigenstate with energy $E_{\lambda_i}$ on the spherical basis.
Fig.\ref{fig15} shows how the inverse participation ratio
$\xi$ (obtained averaging $\xi_{\lambda_i}$ in one--shell intervals
$E(n-1/2)\leq E_{\lambda_i} \leq E(n+1/2)$, with $E(x)\equiv -1/2 x^2$)
increases with the energy for a given static field or equivalently
(in the insert) with the field for a given energy. As a result,
more and more shells are mixed moving to the saddle point 
$\epsilon_{s0}=0.13$ and chaotic properties become stronger. 
For $\epsilon_{s0}=0.02$, $n_0=60$, $\xi\approx 75$
and this indicates that eigenfunctions are significantly spread over
different shells. 

\begin{figure}
\centerline{
\epsfxsize=8cm
\epsfysize=8cm
\epsffile{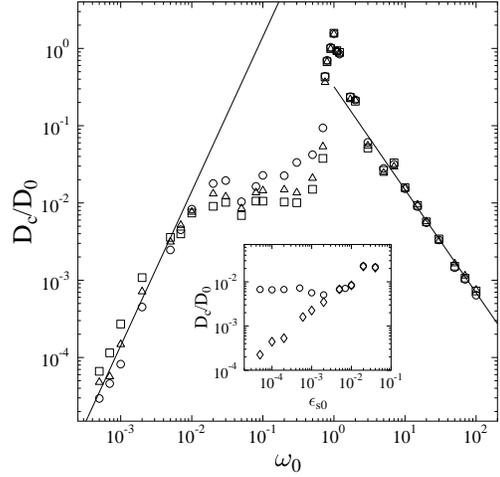}
}
\caption{
Dependence of the scaled classical diffusion rate $D_c/D_0$ on the scaled 
frequency $\omega_0$ for alkali Rydberg atoms in a static 
electric field, at $\epsilon_0=0.005$, 
$\epsilon_{s0}=0.02$, $n_0=60$ for Rb (circles), 
Na (triangles) and Li (squares).
The straight lines show the
theoretical dependence for $\omega_0 \ll 1$ and $\omega_0 \gg 1$,
with $\chi_1=140$, $\chi_2=0.3$. 
Ensembles from $200$ to $1000$ trajectories, initially distributed
microcanonically on the energy shell, have been used.
Core potential parameters for Na as in Fig.\ref{fig13}, for Rb 
$\alpha_1=3.10$, $\alpha_2=1.56$, $\alpha_3=0.76$ and for Li
$\alpha_1=2.84$, $\alpha_2=4.05$, $\alpha_3=4.26$.
The insert shows, for Rb at $\omega_0=0.1$, $n_0=60$, 
the dependence of $D_c/D_0$ on $\epsilon_{s0}$ 
for $\epsilon_0=0.0005$
(diamonds) and the more flat behavior for $\epsilon_0=0.005$
(circles).
}
\label{fig16}
\end{figure}

When the microwave field is turned on, the internal chaos 
leads to diffusive energy excitation in the classical dynamics, with
energy diffusion rate per unit time $D_c=<(\Delta E)^2>/\Delta t$.
As for the hydrogen atom in static electric and microwave fields, the rate
$D_c$ can be compared with the diffusion rate $D_0$ for hydrogen
with only the microwave field present and at $\omega_0=1$.
Again the asymptotic behaviors $D_c=\chi_1 D_0 \omega_0^2$ 
(for $\omega_0\ll 1$) and $D_c=\chi_2 D_0 \omega_0^{-4/3}$
(for $\omega_0\gg 1$) are expected. This is confirmed by 
Fig.\ref{fig16}, that shows the frequency dependence of
the scaled diffusion rate $D_c/D_0$ for Rb, Na and Li at 
$\epsilon_{s0}=0.02$, $\epsilon_0=0.005$, $n_0=60$. The core
potential parameters $\alpha_i$ are not changed
with the initial level $n_0$.
As a result, the dependence on the initial energy or on $n_0$ can 
no longer be eliminated by classical rescaling. However the classical 
scaling rule remains a useful approximation since $D_c/D_0$ varies very 
slowly with $n_0$ for Rydberg atoms (for Rb at 
$\epsilon_{s0}=0.02$, $\epsilon_0=0.005$, $\omega_0=0.1$, $D_c/D_0$ 
remains between $0.019$ and $0.023$ for $20\leq n \leq 80$).
This fact can be understood by taking into account that only for low
$l$ values the electron collides with the core. The energy diffusion
rate is therefore determined by the frequency of these collisions,
which is determined by the frequency of classical precession in
$l$, namely by the scaled Stark frequency $\omega_{s0}=3\epsilon_{s0}$,
independently from $n_0$.
In other words, the dimensions of the core play no important r\^ole
for the classical diffusion rate.

Fig.\ref{fig16} presents some other interesting features: a sharp resonant
peak for $\omega_0=1$ and a plateau for $0.02\leq\omega_0\leq 0.4$.
This latter characteristic is due to the fact that core 
collisions, that are responsible for energy diffusion, occur at
the Stark frequency $\omega_{s0}$, independently from $\omega_0$.
This fact is also illustrated in the insert of Fig.\ref{fig16}: for very
low microwave intensity ($\epsilon_0=0.0005\ll\epsilon_{s0}$) and 
for $\omega_0=0.1$ 
the diffusion coefficient grows with the static
field $\epsilon_{s0}$, while for $\epsilon_0=0.005$,
when $\epsilon_{s0}<\epsilon_0$, the microwave field dominates the 
angular momentum precession and the dependence of $D_c/D_0$ on 
$\epsilon_{s0}$ becomes rather weak.

\begin{figure}
\centerline{
\epsfxsize=8cm
\epsfysize=8cm
\epsffile{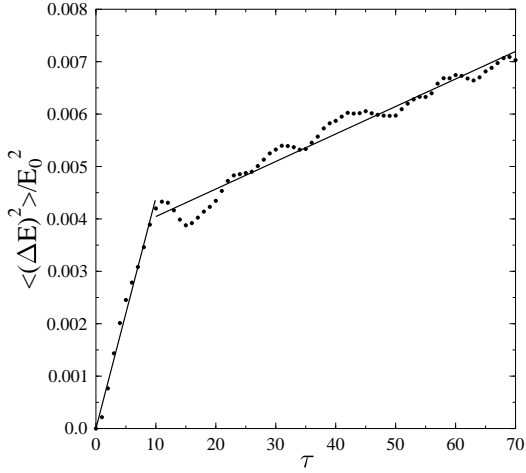}
}
\caption{
Dependence of $\langle(\Delta E)^2\rangle/E_0^2$ on the number of 
microwave periods $\tau$, for $n_0=60$, $\epsilon_{s0}=0.02$,
$\epsilon_0=0.005$, $\omega_0=1$. The straight line fits give
$D_c/D_0=1.4$ (for $0\leq\tau\leq 10$) and $D_c/D_0=0.17$
(for $10\leq\tau\leq 70$).
}
\label{fig17}
\end{figure}

The peak at $\omega_0=1$ in Fig.\ref{fig16} is due to the resonance 
between Kepler and microwave frequency. Fig.\ref{fig17} shows that
the diffusion rate drops abruptly as the electron goes out of
the resonance (for $\epsilon_0=0.005$ this happens after 
approximately $10$ microwave periods). Such sharp change of the diffusion
rate was observed only near the main resonance $\omega_0=1$, while for
other integer values of $\omega_0$ our studies did not indicate any sharp
change. 

\begin{figure}
\centerline{
\epsfxsize=8cm
\epsfysize=8cm
\epsffile{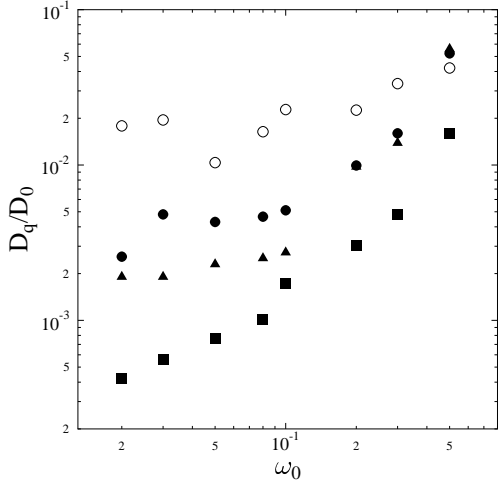}
}
\caption{
Quantum scaled diffusion rate $D_q/D_0$ vs. frequency
$\omega_0$ for Rb (full circles), Na (full triangles) and
Li (full squares) at $\epsilon_{s0}=0.02$, $n_0=60$,
$0.005\leq\epsilon_0\leq 0.03$. Open circles show the 
classical scaled diffusion rate $D_c/D_0$ for Rb,
with parameters $\alpha_i$ as in Fig.\ref{fig16}. 
}
\label{fig18}
\end{figure}

To study diffusion in the quantum case we considered
the initial wavefunction
at $t=0$ as an eigenstate of the Hamiltonian (\ref{hamqdef})
at $\epsilon=0$ with eigenenergy 
$E_{\lambda_0}\approx E_0=-1/2n_0^2$ \cite{qdef}.
As in the case of the hydrogen atom in magnetic and microwave fields,
the quantum dynamics was followed by a split-step method,
similar to the one described in \cite{CCSG87}. 
The quantum diffusion rate $D_q$ was obtained by a linear fit of
the energy square variance $<(\Delta E)^2>$ vs. time for the first 
few microwave periods.
The classical--quantum comparison for the scaled diffusion coefficients
$D_c/D_0$ and $D_q/D_0$ (see Fig.\ref{fig18})
shows a quantitative difference. This is due to
the fact that only low angular momentum values contribute to the
diffusion and so purely quantum effects can be expected. 
In addition, the influence of the ionic core is stronger
in classical mechanics 
because trajectories can penetrate in arbitrarily small regions in
the phase space. On the contrary, quantum mechanics tends to smooth 
over such regions. Apparently this is the reason for which classical
diffusion rate is sistematically larger than the corresponding 
quantum value.

Due to quantum interference effects, 
the above diffusive excitation may eventually stop before ionization.
The resulting quasistationary distribution is characterized
by an exponential decay in the number of absorbed photons,
with a localization length
$\ell_q$ proportional to the one--photon transition rate $\Gamma_q$ and 
to the density of coupled states $\rho_c$: $\ell_q=2\pi\Gamma_q\rho_c$
\cite{Dima87}. The rate is given by 
$\Gamma_q=D_q/\omega^2$ and 
since the internal motion is chaotic the density of coupled states
is $\rho_c=n_0^4$. This leads to \cite{qdef}:
\begin{eqnarray}
\label{locqdef}
\ell_q=\ell_\phi\frac{D_q}{D_0\omega_0^2}n_0,
\end{eqnarray}
where, as in Eq.(\ref{locb}), $\ell_\phi=3.3\epsilon_0^2 n_0^2$
and the conditions $\ell_q>1$, $\omega \rho_c>1$ must be satisfied.

In order to check the theoretical prediction (\ref{locqdef}), the 
quantum evolution of an eigenstate with eigenenergy 
$E_{\lambda_0}\approx -1/2 n_0^2$,
and $n_0=60$ was followed up to $200$ microwave periods
in the Stark basis for $\epsilon_{s0}=0.02$.
The total basis size was up to $1150$ states.
The system parameters were varied in the intervals 
$0.003\leq\epsilon_0\leq 0.03$, $0.02\leq\omega_0\leq 2$, which
corresponds to $15\leq N_I\leq 1500$.

\begin{figure}
\centerline{
\epsfxsize=8cm
\epsfysize=8cm
\epsffile{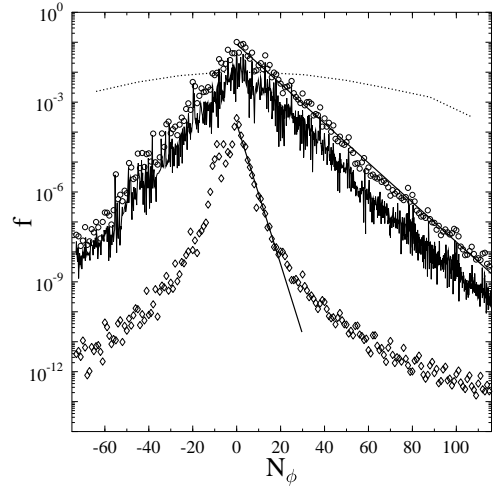}
}
\caption{
Probability distribution vs. number of absorbed photons $N_\phi$:
quantum distribution $f_\lambda$ over the Stark eigenstates 
(full curve), quantum probability in one--photon intervals $f_N$
(circles) and classical probability in one--photon intervals (dotted line)
for Rb at $n_0=60$,
$\epsilon_{s0}=0.02$, $\omega_0=0.08$, $\epsilon_0=0.005$,
$180\leq\tau\leq 200$. Here $D_q/D_0=0.0047$ and the theoretical localization
length (Eq.(\ref{locqdef})) is $\ell_q=13$.
The straight line shows the fit for the exponential localization,
with $\ell_{qN}=13$.
Diamonds give $f_N$ (shifted down by 
$10^3$) for Rb under the same conditions, except for $\epsilon_0=0.002$,
$\ell_q=2.1$. The straight line shows the fit for $0\leq N_\phi\leq 20$
($\ell_{qN}=3.7$), whereas the fit for $N_\phi\geq 50$ gives 
$\ell_{qN}=28$. Classical data are obtained with $\alpha_i$ 
parameters as in Fig.16.
}
\label{fig19}
\end{figure}

Typical examples of stationary distributions for Rb at 
$\omega_0=0.08$ and $n_0=60$ are shown in Fig.\ref{fig19}. 
In order to determine the numerical value $\ell_{qN}$
of the localization length, we first computed the total probability
$f_N$ in each one--photon interval.
Then the least square fit with
with $f_N\sim \exp(-2 N_\phi/\ell_{qN})$ allows to determine
$\ell_{qN}$. In Fig.\ref{fig19} the numerical localization length
$\ell_{qN}$ agrees with
the theoretical $\ell_q$ value (Eq.(\ref{locqdef})) for
$\epsilon_0=0.005$.
The figure also shows how quantum 
excitation is strongly suppressed in comparison with the classical case.
For a smaller microwave intensity ($\epsilon_0=0.002$) the 
fit in the interval
$0\leq N_\phi\leq 20$ gives $\ell_{qN}=3.7$, whereas the fit for
the tail ($N_\phi\geq 50$) shows a much slower decay, with 
$\ell_{qN}=28$. 
We checked that this slope change is not affected by the variation 
of the basis size and integration step. This change in the probability
decay at large $N_\phi$ can be attributed to a significant change in 
the eigenstate structure in a static electric field when one approaches
the saddle point behind which tunneling takes place.
Indeed according to the results shown in Fig.\ref{fig15}
a Stark eigenstates for higly excited levels projects on a larger 
number of hydrogenic levels. Additional investigations are 
required for a better understanding of this effect.  

\begin{figure}
\centerline{
\epsfxsize=8cm
\epsfysize=8cm
\epsffile{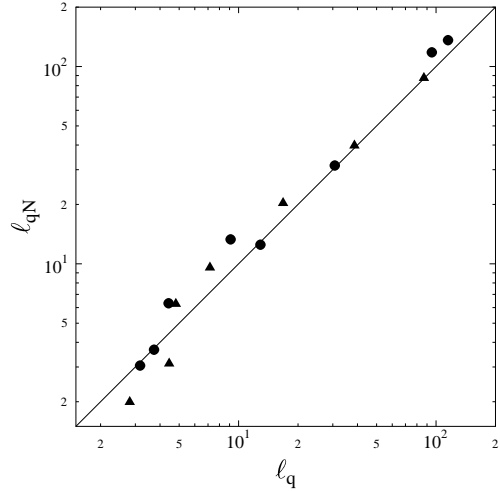}
}
\caption{
The numerically computed localization length $\ell_{qN}$ vs.
the theoretical estimate $\ell_q$ (Eq.(\ref{locqdef}))
for Rb (full circles) and Na
(full triangles) at $n_0=60$, $\epsilon_{s0}=0.02$, $\epsilon_0=0.005$,
$0.02\leq\omega_0\leq 0.5$.
The straight line corresponds to $\ell_{qN}=\ell_q$.
}
\label{fig20}
\end{figure}

The comparison of the numerically obtained localization lengths 
$\ell_{qN}$ vs. the theoretical estimates $\ell_q$ 
confirms the predictions of Eq.(\ref{locqdef}) 
(see Fig.\ref{fig20} for Rb and Na at a fixed $\epsilon_0$ value, with
$\omega_0$ varied over the wide range $0.02\leq\omega_0\leq 0.5$).
We did not study dynamical localization for the Li atom since it is
close to the integrable case.

Equation (\ref{locqdef}) allows to determine the quantum delocalization 
border \cite{qdef} from the condition $\ell_q=N_I$:
\begin{eqnarray}
\label{delocqdef}
\epsilon_q=\frac{1}{n_0}\sqrt{\frac{{D}_0\omega_0}{6.6D_q}}.
\end{eqnarray}
For $\epsilon_0>\epsilon_q$ diffusive ionization takes place. 
Actually, this threshold should be lowered
since ionization is also possible due to the static field 
above the threshold $\epsilon_{s} n_s^4 = 0.13$.
A more accurate estimate is given by the relation
$\ell_q=N_I^s$, with $N_I^s=N_I(1-n_0^2/n_s^2)$. 
Also one should keep in mind that the estimate (\ref{delocqdef})
is based on the initial local value of $D_q/D_0$ taken at $n \approx n_0$.
In the process of excitation the ratio $D_q/D_0$ can be changed,
for example due to a sharp peak near $\omega_0 \approx 1$ 
(see Fig.\ref{fig16}). This can give some additional decrease
for the border $\epsilon_q$.

\begin{figure}
\centerline{
\epsfxsize=8cm
\epsfysize=8cm
\epsffile{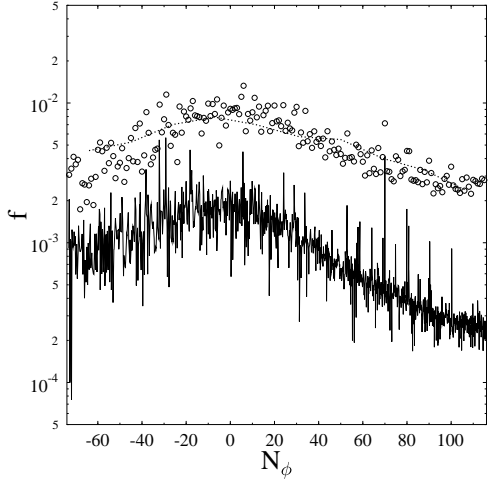}
}
\caption{
Same as in Fig.\ref{fig19} but above the
delocalization threshold $\epsilon_q$ (Eq.(\ref{delocqdef})), 
for Rb at $n_0=60$, 
$\epsilon_{s0}=0.02$, $\omega_0=0.08$, $D_q/D_0=0.0047$,
$\epsilon_0=0.03>\epsilon_q=0.027$, $40\leq \tau\leq 50$. 
}
\label{fig21}
\end{figure}

Fig.\ref{fig21} shows that above the delocalization border
($\epsilon_0>\epsilon_q$) the quantum excitation is close to 
the classical one.
In this case the wave packet escapes into continuum and 
quantum interference effects are unable to freeze quantum 
diffusion before ionization.
We note that the border (\ref{delocqdef}) is lower than for the
hydrogen atom \cite{qdef} approximately by a factor
$\sqrt{n_0}$ which appears due to internal chaos originated
by quantum defects (see also Fig.5 in \cite{qdef}).

The border (\ref{delocqdef}) for alkali Rydgerg atoms 
in static and microwave fields is in qualitative agreement
with a series of experiments by 
Gallagher {\it et al.} \cite{Gall1,Gall2}, that showed at low frequency
($\omega_0\ll 1$) a scaled microwave ionization threshold 
$\epsilon_G\sim 1/n_0$ instead of the static field hydrogenic border
$\epsilon_{s0}=0.13$. Also for Rb atoms the $\epsilon_G$ threshold is
well below the one for hydrogen \cite{Walther}.
Using the data for $D_q$ in the frequency interval 
$0.02\leq\omega_0\leq 0.5$ one can see that the quantum delocalization 
border is approximaely $\epsilon_q\approx 1.5\epsilon_G$
(see Fig.6 in \cite{qdef}).

The ionization process was interpreted by Gallagher {\it et al.}
\cite{Gall1,Gall2} as due to a chain of Landau--Zener
transitions to higher--lying states, until the static
field ionization border is reached. Indeed, in the presence of 
quantum defects,
the first avoided crossing between the $n$ and $n+1$ Stark manifolds
occurs at a field $\epsilon_0=1/3n_0\sim\epsilon_G$. However, this theory
doesn't explain which is the dynamical mechanism that
brings the electron through the whole chain of Stark levels up to ionization.
In addition, the Landau--Zener theory does not apply
if the microwave frequency is larger than the typical spacing between
levels, namely for $\omega_0 n_0>1$.
On the contrary, the dynamical localization theory allows to 
understand ionization of atoms
in a static electric field for the non--adiabatic regime
$\omega_0 n_0>1$.
According to this theory the mechanism of ionization is qualitatively
different from the one proposed by 
Gallagher {\it et al.}, namely ionization takes place
due to diffusive excitation in energy originated by internal chaos
existing in absence of the microwave field.

\begin{figure}
\centerline{
\epsfxsize=8cm
\epsfysize=8cm
\epsffile{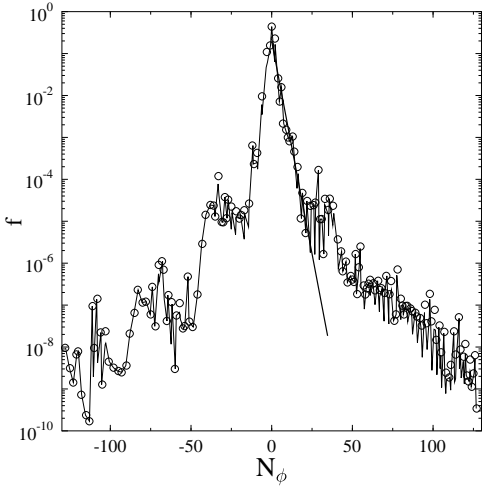}
}
\caption{
Same as in Fig.\ref{fig19}, for the conditions of Fig.2d in
[21], for Na at $n_0=28$,
$\epsilon_{s0}=0.024$, $\omega_0=0.027$, $\epsilon_0=0.002$,
$180\leq\tau\leq 200$,
$D_q/D_0=0.0027$, $\ell_q=1.1$, $\ell_{qN}=4$ (straight line)
for $f_N>10^{-5}$
and $\ell_{qN}=25$ for $f_N<10^{-5}$.
}
\label{fig22}
\end{figure}

A difficulty for the direct comparison with experiments 
\cite{Walther,Gall1,Gall2,Blumel} is
that only few of them were done in the presence of a 
a static electric field \cite{Gall2} and, in addition, they 
were in the regime $\omega\rho_c=\omega_0 n_0<1$. However,
there is a case for Na at $n_0=28$, initial orbital
momentum $l_0=2$, $\epsilon_{s0}=0.024$,
$\omega_0=0.027$ (Fig.2d in \cite{Gall2}) which is not far from
our conditions ($\omega_0 n_0=0.76$).
The experimental $50\%$ ionization threshold is $\epsilon_{0ex}=0.002$.
This value is about $20$ times smaller than the quantum delocalization
border given by Eq.(\ref{delocqdef})
with $D_q/D_0=0.0027$. Indeed, quantum simulations
in the eigenstate basis, extended up to $200$ periods (see Fig.\ref{fig22}),
give a localization length $\ell_{qN}=4\ll N_I=520$. 
We understand this discrepancy as due to deviations in the tail
of probability distribution ($\ell_{qN}=25$ for $f_N<10^{-5}$)
similarly to the case discussed in Fig.\ref{fig19}.

\begin{figure}
\centerline{
\epsfxsize=8cm
\epsfysize=8cm
\epsffile{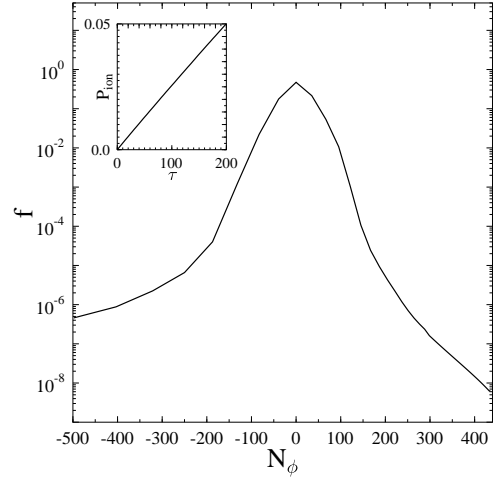}
}
\caption{
Probability distribution over the spherical hydrogenic
basis for the case of Fig.\ref{fig22}.
Time evolution is followed under the same static and 
microwave field conditions, starting from a hydrogenic eigenstate
with $n_0=28$ and initial value of the orbital momentum quantum number 
$l_0=2$, with absorption parameters
(Eq.(\ref{absorption})) given by $\gamma_n=1/2\pi n_s^3$ for 
$\epsilon_s n^4>0.13$ ($n>n_s=42$).
The insert shows that ionization probability depends on time in a 
nearly linear way (for the interaction time here considered, $\tau\leq 200$), 
with ionization rate (per microwave period) 
$\Gamma=2.5\times 10^{-4}$.  
}
\label{fig23}
\end{figure}

In order to support this interpretation, we followed the time evolution
in the hydrogenic basis and simulated tunneling ionization by an absorption
mechanism. Namely, we modified the evolution of the wave function 
$\psi(t)$ as
\begin{eqnarray}
\label{absorption}
\psi(t)=T\exp(-i\hat{H}t)\exp(-\frac{\hat{\gamma}}{2}t)\psi(0),
\end{eqnarray}
where $T$ is time ordering operator,
$\hat{\gamma}$ is a diagonal operator in the spherical basis, with 
matrix elements $\gamma_n\approx 1/2\pi n_s^3$ for all the states
in a shell with principal quantum number $n>n_s$ ($\epsilon_s n_s^4=0.13$).
In this way, there is absorption for levels 
above the static field ionization threshold after a time approximately 
equal to the Kepler period $T_{n_s}=2\pi n_s^3$.
In this model, ionization probability grows with time in a nearly linear 
way (see Fig.\ref{fig23}), 
with ionization rate (per microwave period)
$\Gamma=2.5\times 10^{-4}$.
This fact can become important for very long
interaction times
($\tau_I=8.3\times 10^3$ microwave periods in
\cite{Gall2}) leading to strong ionization.

Another situation in which the experimental ionization border is
much lower than the one given by the theoretical estimate
(\ref{delocqdef}) was observed in Beterov {\it et al.} experiments
\cite{Beterov}. In this experiments, with $n_0=36$, $l_0=1$,
$\omega_0=0.51$, $\epsilon_{s0}=0$, approximately half of Na atoms
were ionized after $\tau_I=1.5\times 10^5$ microwave periods
at $\epsilon_0=0.003$. A static field of strength $\epsilon_{s0}=0.02$
would give a diffusion rate $D_q=0.057 D_0$. In such a case the quantum
delocalization border is $\epsilon_q=0.032$ which is much higher than
the experimental value. The numerical simulations with effective 
absorption discussed above give an absorption rate 
$\Gamma=9.3\times 10^{-7}$ for $\epsilon_0=0.003$ and $\tau\leq 200$.
Such ionization rate would give a significant ionization during the
long interaction time $\tau_I$ used in experiments \cite{Beterov}.

A possibility, alternative to the presence of a weak static electric
field, is that some noise existing in the waveguide could destroy
localization and give a larger ionization probability compared to the 
theoretical explanation (\ref{delocqdef}).

For $\omega_0\ll 1$, one could expect that the slowly varying microwave
field will also play the r\^ole of a static electric field even if 
$\epsilon_s=0$. However, in reality, this expectation can be valid only if
$\omega_0$ is much smaller than the frequency 
$\omega_{s0}$ of classical precession
determined by the Stark splitting and equal to $3\epsilon_0$
($\omega_0\ll 3\epsilon_0$). If this condition is not satisfied
then the mixing of $l$ states will not take place and this is in
agreement with our numerical data for Rb at $n_0=60$, $l_0=0$,
$\omega_0=0.1$, $\epsilon_0=0.01$, $\epsilon_{s0}=0$, where after
$\tau=100$ microwave periods only few $l$--states are mixed
($<l>=2.8$). At the same time in a presence of a static field 
$\epsilon_{s0}=0.02$ the probability spreads over all accessible
$l$--values ($<l>=28$). The localization in $l$ space for Rb atoms 
was found also in \cite{Blumel} at $\epsilon_s=0$.

The existing experiments does not allow unfortunately to make
a direct check of dynamical localization theory because the interaction 
times were too long and therefore it is difficult to control the
effect of environment. However present laboratory conditions allow 
to study short interaction times ($\tau\approx 200$) and high
quantum numbers $n_0\approx 60$. In these conditions, according to our
numerical and theoretical results, quantum excitation of atoms is
well described by the dynamical localization theory. 
Experiments in this regime will
allow to test the quantum localization effects in a range of
parameters much larger than it was so far possible.

Additional investigations should be done for the regime $\omega_0>1$.
In this case the electron precess with very low frequency 
$\omega_{s0}\ll\omega_0$ and therefore the static electric field
gives an adiabatic perturbation on the localized distribution
in the photon number. The case of such adiabatic destruction of 
the localized case was discussed in \cite{CGLSS91} and manifestations
of this effect for alkali Rydberg atoms should be analyzed more
carefully. The situation for alkali Rydberg atoms in a static electric
field is quite different from the case of the hydrogen atom in magnetic 
and microwave fields, where for $\omega_0>1$ the microwave frequency is of
the order of the Larmor frequency ($\omega_{L0}\approx\omega_0$).
On the contrary, for alkali atoms $\omega_{s0}\ll\omega_0$ and therefore
localization takes place faster than the spreading of the wave function
over the whole energy surface.

\section{Conclusions}

In this paper we have analyzed the properties of microwave ionization 
of chaotic Rydberg atoms. Similar to the cases of parity
violation in nuclei \cite{Sushkov-Flambaum} and of two interacting
particles effect in disordered systems \cite{Dima94}, 
a chaotic structure of eigenstates (in absence of microwave) leads
to a chaotic enhancement of radiation interaction with atoms.
As a result the localization length in the number of photons
is strongly increased as 
compared to the usual situation in which the  internal dynamics of the atom,
 without the microwave field, is integrable \cite{CCSG87,IEEE}; as a 
consequence, the quantum delocalization border drops down significantly.
The theory of dynamical localization 
developed for such chaotic atoms is in good agreement with the
results of extensive numerical simulations.

Investigations of such atoms in laboratory experiments
represent a new important opportunity to provide detailed 
results for quantum chaos and dynamical localization.
Indeed, due to internal chaos, the excitation
proceeds in a diffusive way even if the microwave
frequency is much smaller than the Kepler frequency
of electron's rotation ($\omega_0 \ll 1$). As a result
the number of photons $N_I$ required for ionization can be as large as
few thousands, thus  allowing to investigate the dynamical
localization of chaos in great detail.
Experimental conditions are rather similar to those in previous  
experiments \cite{Walther,Kleppner,Gall1,Gall2,Klep2,Beterov} 
and are available now in modern laboratories.

Here  we have discussed the case of parallel fields
in which the magnetic quantum number $L_z$ remains an integral of motion.
It can be also interesting to study a more general situation
with arbitrary field's orientation and polarization.
In this case the density of coupled states will be even larger
$(\rho_c \sim n_0^5)$. This can lead to  an additional growth
of localization length by a factor $n_0$ and to a decrease
of the delocalization border by a factor $\sqrt{n_0}$.
At the same time the effective ``sample'' size $N_I$
can be increased by $n_0$  for such small
frequencies as $\omega_0 \approx 1/n_0^2$.
However, this case deserves more detailed studies. Indeed, the 
existence of additional approximate integrals of motion 
or the appearance of very slow adiabatic frequencies is not
excluded (especially for a static electric field) and 
this may lead to  new interesting results.

The results of  the present paper also show that theoretical and 
experimental studies
of chaotic Rydberg atoms still represent a challenge for 
fundamental research of quantum chaos.

\newpage
\vglue 1.cm
{\centerline {\bf FIGURES}}

\end{multicols}
\end{document}